\documentclass[%
 longbibliography,
 reprint,
 amsmath,amssymb,
 aps,
 pra,
 superscriptaddress
]{revtex4-1}

\usepackage{graphicx}
\usepackage{bm}
\usepackage{hyperref}
\usepackage{braket}
\usepackage[utf8]{inputenc}
\usepackage{qcircuit}
\usepackage{color}
\usepackage{comment}
\usepackage{mathtools}

\newcommand{\mr}[1]{\mathrm{#1}}

\begin{document}

\title{Calculating nonadiabatic couplings and Berry's phase by variational quantum eigensolvers}
\author{Shiro Tamiya}
 \email{tamiya@qi.t.u-toyko.ac.jp}
 \affiliation{Department of Applied Physics, Graduate School of Engineering, The University of Tokyo, 7-3-1 Hongo, Bynkyo-ku, Tokyo 113-8656, Japan}
 \affiliation{Photon Science Center, Graduate School of Engineering, The University of Tokyo, 7-3-1 Hongo, Bunkyo-ku, Tokyo 113-8656, Japan}
 
\author{Sho Koh}
\email{koh@qunasys.com}
 \affiliation{QunaSys Inc., Aqua Hakusan Building 9F, 1-13-7 Hakusan, Bunkyo, Tokyo 113-0001, Japan}

\author{Yuya O. Nakagawa}
 \email{nakagawa@qunasys.com}
 \affiliation{QunaSys Inc., Aqua Hakusan Building 9F, 1-13-7 Hakusan, Bunkyo, Tokyo 113-0001, Japan}
 
\date{\today}

\begin{abstract}
The variational quantum eigensolver (VQE) is an algorithm to find eigenenergies and eigenstates of systems in quantum chemistry and quantum many-body physics. The VQE is one of the most promising applications of near-term quantum devices to investigate such systems. Here we propose an extension of the VQE to calculate the nonadiabatic couplings of molecules in quantum chemical systems and Berry's phase in quantum many-body systems. Both quantities play an important role to understand the properties of a system beyond the naive adiabatic picture, e.g., nonadiabatic dynamics and topological phase of matter. We provide quantum circuits and classical post-processings to calculate the nonadiabatic couplings and Berry's phase. Specifically, we show that the evaluation of the nonadiabatic couplings reduces to that of expectation values of observables while that of Berry's phase also requires one additional Hadamard test. Furthermore, we simulate the photodissociation dynamics of a lithium fluoride molecule using the nonadiabatic couplings evaluated on a real quantum device. Our proposal widens the applicability of the VQE and the possibility of near-term quantum devices to study molecules and quantum many-body systems.
\end{abstract}

\maketitle

\section{Introduction}
Quantum computers currently available or likely to be available in the near future are attracting growing attention. They are referred to as noisy intermediate-scale quantum (NISQ) devices~\cite{Preskill2018}, comprising tens or hundreds of qubits without quantum error correction. While it remains unclear whether they have ``quantum advantage" over classical computers, the fact that they work explicitly based on the principle of quantum mechanics motivates researches on finding applications and developing quantum algorithms for practical problems that are classically intractable~\cite{RevModPhys.92.015003, Cao2019, Mitarai2018, Farhi2018, Havlicek2019, Kusumoto2019article, Farhi2014, Cong_2019, cerezo2020variational, Romero_2017, Sharma_2020, McClean_2016, endo2020hybrid}. In particular, investigating quantum many-body systems with the variational quantum eigensolver (VQE)~\cite{Peruzzo2014} is believed to be one of the most promising applications for NISQ devices.

The VQE is an algorithm to obtain eigenenergies and eigenstates of a given quantum Hamiltonian.
In the VQE, quantum and classical computations are separated appropriately, and interactive quantum-classical hybrid architecture eases the difficulty of implementing the algorithm in the NISQ devices~\cite{Peruzzo2014, OMalley2016, Kandala2017, Colless2018, Hempel2018, Kandala2019}.
The VQE, which was originally proposed for finding the eigenenergy of the ground state, has been extended to find the excited energies and states~\cite{McClean2017PRA, Colless2018, Nakanishi2019PRR, Parrish2019, Jones2019, Higgott2019, PhysRevResearch.2.043140}, non-equilibrium steady states~\cite{PhysRevResearch.2.043289, liu2021variational}, derivatives of eigenenergies with respect to external parameters of the system~\cite{Mitarai2019Derivative, Parrish2019Hybrid,2019npjQI...5..113O}, and the Green's function~\cite{Endo2019Green}.

This study aims to add a new recipe to the catalog of the VQE-based algorithms for quantum systems. We propose a method to calculate the nonadiabatic couplings (NACs)~\cite{doi:https:/doi.org/10.1002/9780470141403.ch1, Yarkony2012} of molecules in quantum chemistry and Berry's phase~\cite{Berry1984, Xiao2010, Cohen2019} of quantum many-body systems by utilizing the results of the VQE.
Both quantities are related to the variation of slow degrees of freedom of the system and play a crucial role in the study of quantum chemistry, condensed matter physics, optics, and nuclear physics~\cite{Tully1990, Tully2012, Tavernelli2015, Takatsuka2015,Nakahara2003, Xiao2010, Cohen2019}.

The NACs in quantum chemistry are defined as couplings between two different electronic states under the Born-Oppenheimer approximation~\cite{Born1927}, which are induced nonadiabatically by motions of nuclei (vibrations).
They are fundamental in the nonadiabatic molecular dynamics simulations to study various interesting dynamical phenomena such as photochemical reactions around the conical intersection and electron transfers~\cite{Tully1990, Tully2012, Tavernelli2015, Takatsuka2015}.
On the other hand, Berry's phase is defined as a phase acquired by an eigenstate when external parameters of a system are varied adiabatically along a closed path in the parameter space.
It reflects intrinsic information about a system such as topological properties of materials.
For example, several symmetry-protected topological phases are characterized by Berry's phase~\cite{Asboth2016, Hatsugai2006, Kariyado2018, Araki2020}.
Berry's phase has become influential increasingly in many fields of modern physics, including condensed matter physics and high-energy physics~\cite{Nakahara2003, Xiao2010, Cohen2019}.

Mathematically, the NACs and Berry's phase are related to derivatives of eigenstates with respect to external parameters of a system. In this study, in order to evaluate the NACs and Berry's phase based on the VQE, we develop analytical formulas and explicit quantum circuits to calculate the inner products related to the derivatives of the eigenstates.
A naive way of calculating the NACs based on the VQE requires the Hadamard test~\cite{1998RSPSA.454..339C} with a lot of controlled operations
In contrast, our proposed methods for the NACs are based on the measurements of expectation values of observables, which is tractable on NISQ devices, and do not require the Hadamard test.
As for Berry's phase, there is a previous study~\cite{Murta2020} to calculate it by simulating adiabatic dynamics and performing the Hadamard test at each time step.
That method cannot avoid the undesired time- and energy-dependent dynamical phase contribution in addition to Berry's phase.
Our proposed method for Berry's phase can remove dynamical phase contribution by utilizing the definition of Berry's phase although it still requires the Hadamard test at most once.
Finally, as a demonstration of our methods, we present the simulation of photodissociation dynamics of a lithium fluoride molecule with the value of the nonadiabatic couplings evaluated on the real quantum device, IBM Q Experience~\cite{IBMQ2020}, by our proposed methods.
Our results enlarge the possible scopes of the VQE algorithm and the NISQ devices for simulating various quantum systems.

The rest of the paper is organized as follows.
We briefly review the definition of the NACs and Berry's phase in Sec.~\ref{sec: review of NDC}.
The VQE algorithm is also reviewed in Sec.~\ref{sec: review of VQE}.
Our main results are presented in Secs.~\ref{sec: NAC method} and \ref{sec: Berry method}, where we describe the ways to calculate the NACs and Berry's phase based on the VQE.
The results of the experiment of estimating the nonadiabatic coupling using IBM Q hardware and the simulation of photodissociation dynamics with our methods are shown in Sec.~\ref{sec: experiment}.
The discussion about the cost analysis for running our algorithms on quantum devices is provided in Sec.~\ref{sec: disussion}.
We conclude our study in Sec.~\ref{sec: conclusion}.
Appendices provide details of the experiments, mathematical proofs of the cost analysis, and
further numerical demonstrations of our algorithms.

\section{Review of the nonadiabatic couplings and Berry's phase \label{sec: review of NDC}}
In this section, we review definitions of the NACs~\cite{doi:https:/doi.org/10.1002/9780470141403.ch1, Yarkony2012} and Berry's phase~\cite{Berry1984}. 
Let us consider a quantum system which has external parameters $\vec{R}=(R_1,\ldots,R_{N_x})\in \mathbb{R}^{N_x}$.
These parameters $\vec{R}$ characterize the system, e.g., coordinates of nuclei in the case of quantum chemistry, electromagnetic field applied to a system in the case of conducting metals.
We call $\vec{R}$ as ``system-parameters" and represent the Hamiltonian of the system which depends on $\vec{R}$ by $H(\vec{R})$.
The eigenvalues and eigenstates of $H(\vec{R})$ are denoted by $\{E_i(\vec{R})\}_i$ and $\{\ket{\chi_i(\vec{R})}\}_i$.
We assume that $\{E_i(\vec{R})\}_i$ and $\{\ket{\chi_i(\vec{R})}\}_i$ depend on $\vec{R}$ smoothly and that there is no degeneracy in the eigenspectrum unless explicitly stated in the text.
When there is a degeneracy in the spectrum, the NACs is not well-defined among degenerate eigenstates. Berry's phase is generalized to non-abelian one, i.e., SU(N) matrix for $N$-degenerate ground states~\cite{Nakahara2003}, and the components of the matrix can be determined in a similar way for abelian Berry's phase for the non-degenerate ground state studied in this paper.

\subsection{Nonadiabatic couplings}
Here let us consider a molecular system and $H(\vec{R})$ as the electronic Hamiltonian. Definitions of the first-order NAC (1-NAC) $d_{kl}^I$ and the second-order NAC (2-NAC) $D_{kl}^I$ are as follows,
\begin{eqnarray}
  d_{kl}^I(\vec{R}) &= \braket{\chi_k(\vec{R})|\frac{\partial}{\partial R_I}|\chi_l(\vec{R})}, \label{eq: 1-NAC def} \\
  D_{kl}^I(\vec{R}) &= -\braket{\chi_k(\vec{R})|\frac{\partial^2}{\partial R_I^2}|\chi_l(\vec{R})}, \label{eq: 2-NAC def}
\end{eqnarray}
where $k$ and $l$ are different indices for eigenlevels and $I=1,\ldots,N_x$ denotes the index for the system-parameters.
The Hellman-Feynman theorem~\cite{1933ZPhy...85..180H, Feynman1939} gives a simpler expression of the 1-NAC as 
\begin{equation}
    d_{kl}^I =- \frac{\braket{\chi_k(\vec{R})|\frac{\partial H}{\partial R_I}|\chi_l(\vec{R})}}{E_k(\vec{R})-E_l(\vec{R})}, \label{eq: 1-NAC formula}
\end{equation}
which means that the 1-NAC becomes large when two eigenstates are close to degenerate ($E_k\sim E_l$).
We take advantage of this expression when calculating the 1-NAC in Sec.~\ref{sec: NAC method}.
The 1-NAC lies in the heart of various nonadiabatic molecular dynamics algorithms such as the Tully's fewest switches method~\cite{Tully1990, Tully2012} and {\it ab initio} multiple spawning~\cite{Benkun2000,Benkun2002}.

Equation~\eqref{eq: 2-NAC def} in the case of $k=l$ is related to the diagonal Born-Oppenheimer correction (DBOC) defined as
\begin{equation}
 \begin{split}
  E_\mr{DBOC}&(k)= D_{kk}^{I}(\vec{R}) \\
  &=  -\sum_{\substack{m \\ \alpha=x,y,z}} \frac{1}{2M_m}  \braket{\chi_k(\vec{R})|\frac{\partial^2}{\partial R_{m_\alpha}^2}|\chi_k(\vec{R})} \label{eq: DBOC formula},
 \end{split}
\end{equation}
where $k$ is the eigenlevel to be considered, $M_m$ is the mass of the nucleus $m$, and $R_{m_\alpha}$ is $\alpha$-cordinate ($\alpha=x,y,z$) of the nucleus $m$. It is argued that this correction sometimes brings out crucial differences in stability and dynamics of molecules~\cite{Handy1986, Valeev2003, Ryabinkin2014, Gherib2016}.

In addition, we comment on the gauge invariance of the NACs.
Overall phase factors of eigenstates are arbitrary in general, so there is a $U(1)^M$ degree of freedom in the definition of the NACs,
\begin{equation}
 \ket{\chi_k(\vec{R})} \to e^{i\Theta_k(\vec{R})} \ket{\chi_k(\vec{R})},
 \label{eq: gauge transformation}
\end{equation}
where $k=0, \ldots, M-1$, $M$ is the number of eigenlevels to be considered, and $\Theta_k(\vec{R}) \in \mathbb{R}$ is an arbitrary smooth function of $\vec{R}$.  
The 1-NAC (Eq.~\eqref{eq: 1-NAC def}) and the 2-NAC (Eq.~\eqref{eq: 2-NAC def}) are not invariant under the transformation~\eqref{eq: gauge transformation}.
This dependence
 must be resolved in each algorithm utilizing the value of the NACs.
For example, see Refs.~\cite{Errea2004, Vibok2005, Miao2019}.
We note that real-valued eigenfunctions are usually considered in quantum chemistry, but complex eigenfunctions may be obtained in the VQE in general.

\subsection{Berry's phase}
Berry's phase~\cite{Berry1984} is defined for a closed loop $\mathcal{C}$ in the parameter space $\mathbb{R}^{N_x}$ as,
\begin{eqnarray}
  \Pi_\mathcal{C} = - i \int_\mathcal{C} d\vec{R} \cdot \braket{\chi_k(\vec{R})|\frac{d}{d\vec{R}}|\chi_k(\vec{R})},
  \label{eq: def Berry}
\end{eqnarray}
where $\int_\mathcal{C} \ldots$ is the line integral along the closed loop $\mathcal{C}$, $\ket{\chi_k(\vec{R})}$ is the $k$-th eigenstate of the Hamiltonian $H(\vec{R})$.
If one prepares the $k$-th eigenstate of the system $\ket{\chi_k(\vec{R}_0)}$ at some system-parameters $\vec{R}_0$ and adiabatically varies them in time along $\mathcal{C}$,
the final state will obtain the phase $e^{-i \Pi_\mathcal{C}}$ in addition to the dynamical phase.
We note that Berry's phase is always real by definition because the normalization condition $\braket{\chi_k(\vec{R})|\chi_k(\vec{R})}=1 $ leads to  $\frac{d}{d\vec{R}}(\braket{\chi_k(\vec{R})|\chi_k(\vec{R})}) = 2 \mathrm{Re}\left(\braket{\chi_k(\vec{R})|\frac{d}{d\vec{R}}|\chi_k(\vec{R})}\right)=\vec{0}$.

Finally, we point out the gauge invariance of Berry's phase. The eigenstates have $U(1)$ gauge freedom stemming from arbitrariness of overall phases for them. Under $U(1)$ gauge transformation (Eq.~\eqref{eq: gauge transformation}), Berry's phase is invariant only up to an integer multiple of $2\pi$. 
Since Berry's phase appears as $e^{-i\Pi_\mathcal{C}}$, this arbitrariness does not affect the physics, and we can consider Berry's phase as an observable property of the system~\cite{Nakahara2003, Xiao2010, Cohen2019}.

\section{Review of Variational Quantum Eigensolver \label{sec: review of VQE}}
In this section, we review the VQE algorithm~\cite{Peruzzo2014} to obtain a ground state and excited states of a given Hamiltonian.
We also describe how to compute analytical derivatives of optimal circuit-parameters of the VQE with respect to system-parameters of the Hamiltonian.
Methods described in this section are repeatedly used in Secs.~\ref{sec: NAC method} and \ref{sec: Berry method} to calculate the 1- and 2-NACs and Berry's phase.

Again, let us consider an $n$-qubit quantum system whose Hamiltonian is $H(\vec{R})$.
In the VQE, we introduce an ansatz quantum circuit $U(\vec{\theta})$ and the ansatz state $\ket{\varphi_0(\vec{\theta})}$ in the form of
\begin{equation}
  \ket{\varphi_0(\vec{\theta})} = U(\vec{\theta}) \ket{{\psi_0}},
\end{equation}
where $\ket{\varphi_0}$ is a reference state and $\vec{\theta} = (\theta_1,\ldots,\theta_{N_\theta}) \in \mathbb{R}^{N_\theta}$ is a vector of circuit-parameters contained in the ansatz circuit. We assume $U(\vec{\theta})$ to be a product of unitary matrices each with one parameter,
\begin{equation}
    U(\vec{\theta}) = U_{N}(\theta_{N})\cdots U_2(\theta_2)U_1(\theta_1).
\end{equation}
We also assume each unitary $U_a(\theta_a)$ consists of non-parametric quantum gates and parametric gates in the form of $e^{ig_a P_a\theta_a}$ generated by a Pauli product $P_a\in \{I, X, Y, Z\}^{\otimes n}$ with a coefficient $g_a\in \mathbb{R}$ ($a=1,\ldots,N_\theta$).
Note that many ans{\"a}tze proposed in previous studies fall into this category~\cite{Peruzzo2014, Kandala2017, Gard2020, Parrish2019, Lee2019, Grimsley2019, 2021PRXQ....2b0310T, Matsuzawa2020}.
We will represent $U_j(\theta_j)\cdots U_i(\theta_i)$ as $U_{i:j}$ for simplicity.

\subsection{Variational quantum eigensolver for ground state and excited states}
The original VQE algorithm finds a ground state of a given Hamiltonian based on the variational principle of quantum mechanics.
In the VQE, one optimizes the circuit-parameters $\vec{\theta}$ variationally by classical computers so that the expectation value
\begin{equation}
 E_0(\vec{\theta}, \vec{R}) = \braket{\varphi_0(\vec{\theta})|H(\vec{R})|\varphi_0(\vec{\theta})}
\end{equation}
is minimized with respect to $\vec{\theta}$.
When the ansatz circuit has sufficient capability of expressing the ground state of $H(\vec{R})$ and the circuit-parameters $\vec{\theta}$ converge to optimal ones $\vec{\theta}^*$, we can expect the optimal state $\ket{\varphi_0(\vec{\theta}^*)}$ will be a good approximation to the ground state.
Since tasks of evaluation and optimization of quantum circuits are distributed to quantum and classical computers, it is easier to implement the algorithm on the near-quantum devices~\cite{Peruzzo2014, OMalley2016, Kandala2017, Colless2018, Hempel2018, Kandala2019}.

After the proposal of the original VQE algorithm, there are a variety of extensions of the VQE to find excited states of a given Hamiltonian~\cite{McClean2017PRA, Colless2018, Nakanishi2019PRR, Parrish2019, Jones2019, Higgott2019, PhysRevResearch.2.043140}.
As we will see in Secs.~\ref{sec: NAC method} and \ref{sec: Berry method}, one has to compute (approximate) eigenenergies and transition amplitudes of several Pauli operators between obtained eigenstates to calculate the NACs.
From this viewpoint, the most appropriate methods to calculate them are the subspace-search VQE (SSVQE)~\cite{Nakanishi2019PRR} algorithm and its cousin algorithm, the multistate contracted VQE (MCVQE) algorithm~\cite{Parrish2019}.
Here we briefly describe the SSVQE just for completeness, but formulas for the MCVQE are quite similar.

To obtain approximate eigenenergies and eigenstates up to $i=0,\ldots,M-1$, the SSVQE algorithm uses $M$ easy-to-prepare orthonormal states $\{\ket{\psi_i}\}_{i=0}^{M-1}$ (e.g. computational basis) as reference states.
For our algorithms to work, the reference states also have to be chosen so that we can readily prepare the superpositions of them on quantum devices.
The SSVQE proceeds so as to minimize the following cost function,
\begin{equation}
 \mathcal{L}_{\vec{R}}(\vec{\theta})=\sum_{i=0}^{M-1}w_i\bra{\psi_i}U^{\dag}(\vec{\theta}) H(\vec{R}) U(\vec{\theta})\ket{\psi_i}, \label{eq: cost of SSVQE}
\end{equation}
where $\{w_i\}_{i=0}^{M-1}$ are positive and real weights which satisfy $w_0>w_1>\cdots>w_{M-1}>0$.
When the cost function converges to the minimum at $\vec{\theta}^*(\vec{R})$, it follows that
\begin{align}
 \ket{\varphi_i(\vec{R})} &= U(\vec{\theta}^*(\vec{R}))\ket{\psi_i}, \\
 \tilde{E}_i(\vec{R}) = &\braket{\varphi_i(\vec{R})|H(\vec{R})|\varphi_i(\vec{R})}, 
\end{align}
are good approximations of the eigenstates and eigenenergies, respectively.

One of the most distinctive features of the SSVQE and the MCVQE algorithms is that one can readily compute transition amplitudes $\braket{\varphi_k(\vec{R})|A|\varphi_l(\vec{R})}$ of any observable $A$ between the (approximate) eigenstates obtained. 
Although evaluation of the transition amplitude between two quantum states requires the Hadamard test in general, which contains a lot of extra and costly controlled gates ~\cite{Mitarai2019Methodology}, the SSVQE and the MCVQE circumvent the difficulty by preparing superposition of two eigenstates. 
It is possible to evaluate the transition amplitude by low-cost quantum circuits without extra controlled gates as
\begin{equation}
\begin{split}
 \mathrm{Re}& \left( \braket{\varphi_k(\vec{R})|A|\varphi_l(\vec{R})} \right)\\
 & = \frac{1}{2} \left( \braket{\varphi_{k,l}^{+}(\vec{R})|A|\varphi_{k,l}^{+}(\vec{R})}
  - \braket{\varphi_{k,l}^{-}(\vec{R})|A|\varphi_{k,l}^{-}(\vec{R})} \right), \\
 \mathrm{Im}& \left(\braket{\varphi_k(\vec{R})|A|\varphi_l(R)} \right) \\
 & = -\frac{1}{2} \left( \braket{\varphi_{k,l}^{i+}(\vec{R})|A|\varphi_{k,l}^{i+}(\vec{R})}
 - \braket{\varphi_{k,l}^{i-}(\vec{R})|A|\varphi_{k,l}^{i-}(\vec{R})} \right),
\label{eq: trans amp}
\end{split}
\end{equation}
where $\ket{\varphi_{k,l}^{\pm}(\vec{R})} = U(\vec{\theta}^*(\vec{R}))(\ket{\psi_k}\pm\ket{\psi_l})/\sqrt{2}$ and $\ket{\varphi_{k,l}^{i\pm}(\vec{R})}=U(\vec{\theta}^*(\vec{R}))(\ket{\psi_k}\pm i\ket{\psi_l})/\sqrt{2}$.
Since each term of the right hand sides of the equation is an expectation value of the observable, the evaluation of the transition amplitude is tractable on near-term quantum devices.

\subsection{Derivatives of optimal parameters}
To calculate the NACs with the result of the VQE on near-term quantum devices, we also need derivatives of the optimal circuit-parameters $\vec{\theta}^*(\vec{R})$ with respect to the system parameters $\vec{R}$.
These derivatives are given by solving equations~\cite{Mitarai2019Derivative}
\begin{align}
  \sum_{b=1}^{N_\theta}
  \frac{\partial^2 E_0(\vec{\theta}^*(\vec{R}),\vec{R})}{\partial \theta_a \partial \theta_b}
  \frac{\partial \theta^*_b(\vec{R})}{\partial R_I}
 = - \frac{\partial^2 E_0(\vec{\theta}^*(\vec{R}), \vec{R})}{\partial \theta_a \partial R_I}, \label{eq: theta_first_derivative} \\
 \sum_{b=1}^{N_\theta}
  \frac{\partial^2 E_0(\vec{\theta}^*(\vec{R}),\vec{R})}{\partial \theta_a \partial \theta_b}
  \frac{\partial^2 \theta^*_b(\vec{R})}{\partial R_I \partial R_J} = -\gamma^{(IJ)}_a, \label{eq: theta_second_derivative}
\end{align}
where
\begin{align}
    \gamma_c^{(IJ)} &= \sum_{a,b}
    \frac{\partial^3 E_0(\vec{\theta}^*(\vec{R}),\vec{R}) }{\partial\theta_c\partial\theta_a\partial\theta_b}
    \frac{\partial \theta^*_a}{\partial R_I}
    \frac{\partial \theta^*_b}{\partial R_J} \nonumber \\
    &\quad+ 2\sum_{a}
    \frac{\partial^3 E_0(\vec{\theta}^*(\vec{R}),\vec{R})}{\partial\theta_c \partial\theta_a \partial R_J}\frac{\partial \theta^*_a}{\partial R_I}
    +\frac{\partial^3 E_0(\vec{\theta}^*(\vec{R}),\vec{R})}{\partial\theta_c \partial R_I \partial R_J},
\end{align}
simultaneously for $a=1,\ldots,N_\theta$ (with $I,J=1,\ldots,N_x$ fixed).
Now we use notations as follows:
\begin{equation}
\frac{\partial^2 E_0(\vec{\theta}^*(\vec{R}),\vec{R})}{\partial \theta_a \partial R_I} :=
\left. \frac{\partial^2 E_0(\vec{\theta},\vec{R})}{\partial \theta_a \partial R_I} \right|_{\vec{\theta}=\vec{\theta}^*(\vec{R}), \vec{R}=\vec{R}}.
\end{equation}
These formulas (Eqs.\eqref{eq: theta_first_derivative} and \eqref{eq: theta_second_derivative}) can be derived by taking the derivative of $\frac{\partial E_0(\vec{\theta}^{*}(\vec{R}),\vec{R})}{\partial \theta_a}=0$ with respect to $\vec{R}$. For detailed derivation, see Appendix.~A in Ref.~\cite{Mitarai2019Derivative}.
The quantities appearing in Eq.~\eqref{eq: theta_first_derivative} and Eq.~\eqref{eq: theta_second_derivative}, such as $\frac{\partial^2 E_0(\vec{\theta}^*(\vec{R}),\vec{R})}{\partial \theta_a \partial \theta_b}$ and $\frac{\partial^2 E_0(\theta^*(\vec{R}), \vec{R})}{\partial \theta_a \partial R_I}$, can be evaluated quantum circuits on quantum devices using the method shown in Ref.~\cite{Mitarai2019Derivative}.
Therefore one can solve Eq.~\eqref{eq: theta_first_derivative} and Eq.~\eqref{eq: theta_second_derivative} on classical computers and obtain the derivatives of the optimal circuit-parameters $\{ \frac{\partial \theta^*_a(\vec{R})}{\partial R_I}, \frac{\partial \theta^*_a(\vec{R})}{\partial R_I\partial R_J} \}_{a=1}^{N_\theta}$.

\section{Calculating nonadiabatic couplings with variational quantum eigensolver \label{sec: NAC method}}
In this section, we explain how to calculate the 1-NAC and 2-NAC with the VQE.

\subsection{First-order nonadiabatic coupling}
Evaluation of the 1-NAC based on the VQE is simple by utilizing the formula~\eqref{eq: 1-NAC formula}.
First, we perform the SSVQE or the MCVQE and obtain approximate eigenstates $\ket{\varphi_i(\vec{R})}$ and eigenenergies $\tilde{E}_i$ of $H(\vec{R})$.
Then we calculate the derivative of the Hamiltonian, $\frac{\partial H}{\partial R_I}$, on classical computers. 
Specifically, when we use the Hartree-Fock orbitals to construct the second-quantized Hamiltonian, the derivative $\frac{\partial H}{\partial R_I}$ (more precisely, the derivatives of the one- and two-electron integrals in the molecular orbital basis) can be obtained by solving the coupled perturbed Hartree-Fock (CPHF) equation~\cite{Mitarai2019Derivative, Parrish2019Hybrid, 2019npjQI...5..113O}.
The solution of the CPHF equation can be obtained by the standard softwares for quantum chemistry.

Finally, evaluating the transition amplitude $\braket{\varphi_k(\vec{R})|\frac{\partial H}{\partial R_I}|\varphi_l(\vec{R})}$ on quantum devices by using the method of Eq.~\eqref{eq: trans amp} and substituting it into Eq.~\eqref{eq: 1-NAC formula} gives the value of the 1-NAC.

\subsection{Second-order nonadiabatic coupling}
Next, we introduce an analytical evaluation method of the 2-NAC on near-term quantum devices.
After obtaining approximate eigenstates $\{\ket{\varphi_i(\vec{R})}\}_i$ by the SSVQE or the MCVQE, putting them into Eq.~\eqref{eq: 2-NAC def} yields
\begin{equation}
\begin{split}
 & \bra{\varphi_k(\vec{R})}\frac{\partial^2}{\partial R_I^2}\ket{\varphi_l(\vec{R})}  \\
&= \sum_{a,b} \frac{\partial\theta^*_a}{\partial R_I}
\frac{\partial\theta^*_b}{\partial R_I} \braket{\varphi_k|\partial_a\partial_b\varphi_l}+
\sum_{c}\frac{\partial^2\theta^*_c}{\partial R^2_I} \braket{\varphi_k|\partial_c\varphi_l}, \label{eq: 2-NAC VQE}
\end{split}
\end{equation}
where we denote $\frac{\partial}{\partial \theta_a} \frac{\partial}{\partial \theta_b} \ket{\varphi_j}$ and $\frac{\partial}{\partial \theta_c}\ket{\varphi_j}$ as $\ket{\partial_a\partial_b\varphi_j}$ and $\ket{\partial_c\varphi_j}$, respectively.
We note that plugging Eq.~\eqref{eq: 2-NAC VQE} when $k=l$ into Eq.~\eqref{eq: DBOC formula} gives the formula of the DBOC based on the VQE. 

The derivatives of the optimal circuit-parameters such as $\frac{\partial\theta^*_a}{\partial R_I}$ and $\frac{\partial^2\theta^*_c}{\partial R_I^2}$ can be calculated by the method reviewed in Sec.~\ref{sec: review of VQE}.
The terms $\braket{\varphi_k|\partial_a\partial_b\varphi_l}$ and $\braket{\varphi_k|\partial_c\varphi_l}$ can be evaluated with the Hadamard test~\cite{1998RSPSA.454..339C} in a naive way, but its implementation is costly for near-term quantum devices.
Therefore, in the following, we describe how to reduce the evaluation of $\braket{\varphi_k|\partial_a\partial_b\varphi_l}$ and $\braket{\varphi_k|\partial_c\varphi_l}$ to the measurements of the expectation value of observables, which is the standard process of the near-term quantum algorithms.

\begin{figure}
 \includegraphics[width=9cm]{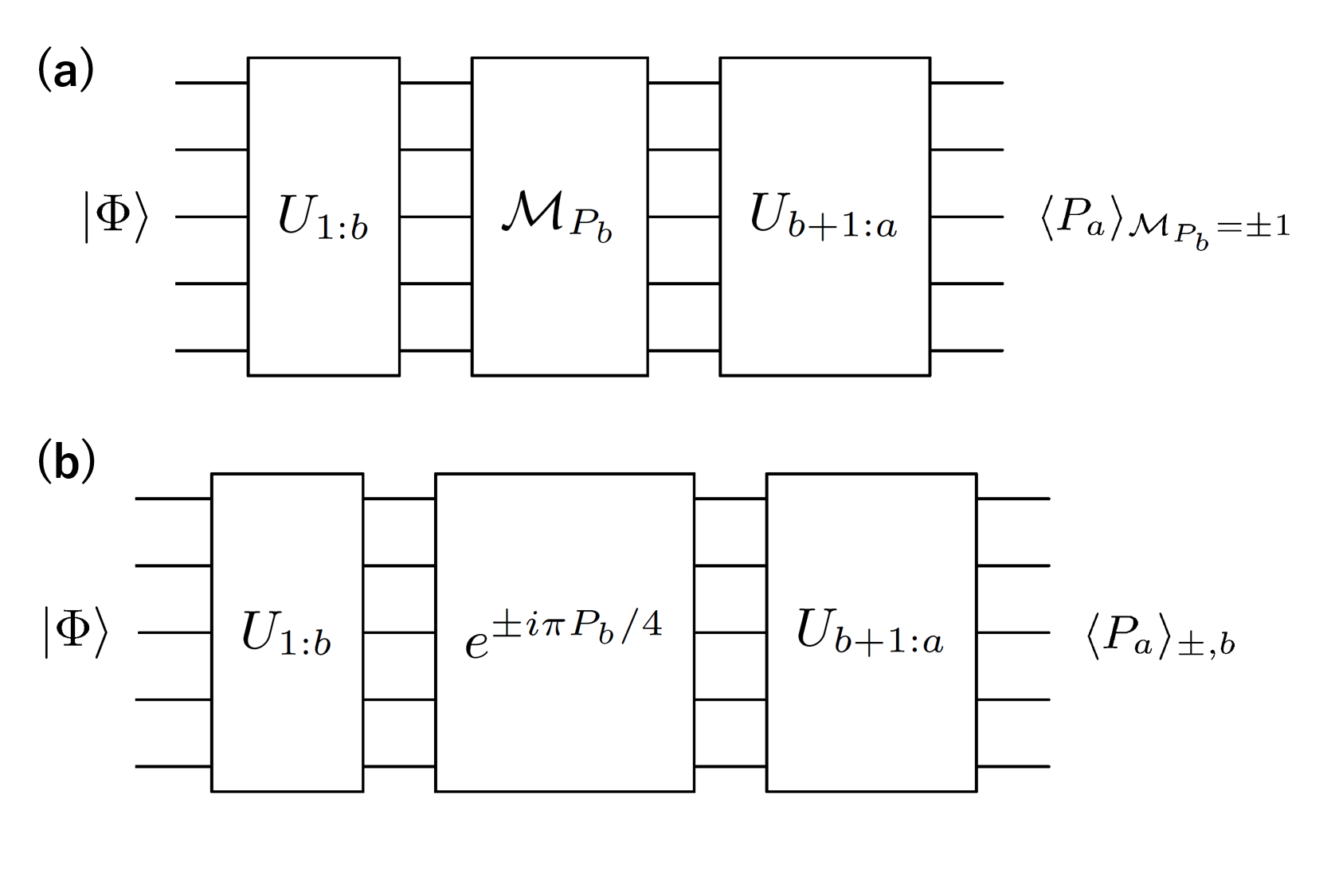}
\caption{(a) Quantum circuit to evaluate Eq.~\eqref{eq: real part} and (b) quantum circuit to evaluate Eq.~\eqref{eq: imag part}. These figures are based on Ref.~\cite{Mitarai2019Methodology}. $\mathcal{M}_{P_b}$ is a projective measurement of the Pauli operator $P_b$.
\label{fig: circuits} }
\end{figure}

\subsubsection{Evaluation of $\braket{\varphi_k|\partial_a\partial_b\varphi_l}$}

To calculate $\braket{\varphi_k|\partial_a\partial_b\varphi_l}$, let us first consider evaluating
\begin{equation}
\bra{\Phi} U^\dag(\vec{\theta}) \frac{\partial}{\partial \theta_a} \frac{\partial}{\partial \theta_b} U(\vec{\theta}) \ket{\Phi} \label{eq: PhideldelPhi}
\end{equation}
with $\ket{\Phi}$ being an arbitrary reference state.
When $a=b$, it follows $\bra{\Phi} U^\dag(\vec{\theta}) \frac{\partial^2}{\partial \theta_a^2}  U(\vec{\theta}) \ket{\Phi}
= -g_a^2 \bra{\Phi} U^\dag(\vec{\theta})U(\vec{\theta}) \ket{\Phi} = -g_a^2$.
When $a \neq b$, we assume $1\leq b < a \leq N_\theta$ without loss of generality.
By using the method in Ref.~\cite{Mitarai2019Methodology}, the real and imaginary parts of Eq.~\eqref{eq: PhideldelPhi} are evaluated separately in the following way.

The real part is calculated with quantum circuits containing projective measurements of the Pauli operator $P_b$ denoted by $\mathcal{M}_{P_b}$,  
\begin{equation}
\begin{split}
    &\mathrm{Re} \left(
    \bra{\Phi} U^\dag(\vec{\theta}) \frac{\partial}{\partial \theta_a} \frac{\partial}{\partial \theta_b} U(\vec{\theta}) \ket{\Phi} \right) \\
    &= -g_a g_b \times \\
    & \left( p(\mathcal{M}_{P_b}=1) \braket{P_a}_{\mathcal{M}_{P_b}=1} - p(\mathcal{M}_{P_b}=-1) \braket{P_a}_{\mathcal{M}_{P_b}=-1} \right),
\end{split}
\label{eq: real PhideldelPhi}
\end{equation}
where
\begin{equation}
\begin{split}
&\braket{P_a}_{\mathcal{M}_{P_b}=\pm1}\\
    &=\frac{1}{4} \frac{\bra{\Phi} U_{1:b}^{\dag} (I\pm P_b) U_{b+1:a}^{\dag} P_a U_{b+1:a} (I\pm P_b) U_{1:b} \ket{\Phi}}{p(\mathcal{M}_{P_b}=\pm1)}    
\end{split}
\label{eq: real part}
\end{equation}
is the conditional expectation value of $P_a$ when the projective measurement of $P_b$ yields $\pm 1$, and 
\begin{equation}
p(\mathcal{M}_{P_b}=\pm1) = \left|\frac{1}{2}(I\pm P_b) U_{1:b}\ket{\Phi}\right|^2 
\end{equation}
is the probability of getting the result $\pm 1$ for the projective measurement of $P_b$.
If $P_b$ is a single Pauli operator or even if $P_b$ is a multi-qubit Pauli operator, we expect that the projective measurement of it can be performed in near-term quantum devices~\footnote{The projective measurement of $P_b$ can be performed by applying a unitary gate $V$ which satisfies $V^\dag P_b V = Z_0$, executing the projective measurement of $Z_0$ and finally applying $V^\dag$ after the projective measurement~\cite{Mitarai2019Methodology}.
Such unitary $V$ can be constructed with $O(\log_2 n )$ depth,  where $n$ is the number of qubits.
First, we transform the non-identity part of $P_b$ into a product of $Z$ gates by using $H$ gates and $S^\dag H$ gates (note that $H\cdot X\cdot H = Z, HS^\dag\cdot Y \cdot HS=Z$). Then CNOT gates are applied to make $Z_i Z_j$ into $Z_i$ by using the equality $\mr{CNOT}_{j,i} \cdot Z_iZ_j \cdot  \mr{CNOT}_{j,i} = Z_i$.
Therefore, the depth of quantum gates needed is at most $1 + 2\log_2 n$.}.
The total circuit for evaluating Eq.~\eqref{eq: real part} is shown in Fig.~\ref{fig: circuits}(a).

On the other hand, the imaginary part of Eq.~\eqref{eq: PhideldelPhi} can be calculated as
\begin{equation}
\begin{split}
\mathrm{Im} &\left( \bra{\Phi} U^\dag(\vec{\theta}) \frac{\partial}{\partial \theta_a} \frac{\partial}{\partial \theta_b} U(\vec{\theta}) \ket{\Phi} \right) \\
 &= \frac{g_a g_b}{2}(\braket{P_a}_{+, b} - \braket{P_a}_{-, b}),
\end{split} 
 \label{eq: imag PhideldelPhi}
\end{equation}
where
\begin{eqnarray}
 && \braket{P_a}_{\pm, b} \nonumber \\
 &=& \bra{\Phi} U_{1:b}^{\dag} e^{\mp i\pi P_b/4} U_{b+1:a}^{\dag} P_a U_{b+1:a} e^{\pm i\pi P_b/4} U_{1:b} \ket{\Phi},
\label{eq: imag part}
\end{eqnarray}
is the expectation value of $P_a$ for the quantum state $U_{b+1:a} e^{\pm i\pi P_b/4} U_{1:b} \ket{\Phi}$.
The circuit for calculation is shown in Fig~\ref{fig: circuits}(b).

Then, to obtain $\braket{\varphi_{k}|\partial_a\partial_b\varphi_{l}}$ we take advantage of the following equality
\begin{equation}
\begin{split}
2\braket{\varphi_{k}|\partial_a\partial_b\varphi_{l}} 
= &\braket{\varphi_{k,l}^+ |\partial_a\partial_b \varphi_{k,l}^+}
-\braket{\varphi_{k,l}^- |\partial_a\partial_b \varphi_{k,l}^-} \\ 
& +\frac{1}{i} \left( \braket{\varphi_{k,l}^{i+} |\partial_a\partial_b \varphi_{k,l}^{i+}}
- \braket{\varphi_{k,l}^{i-} |\partial_a\partial_b \varphi_{k,l}^{i-}} \right),
\end{split}
\label{eq: phik phil}
\end{equation}
where $\ket{\varphi_{k,l}^\pm}=U_{1:N}(\ket{\psi_k}\pm\ket{\psi_l})/\sqrt{2}$ and $\ket{\varphi_{k,l}^{i\pm}}=U_{1:N}(\ket{\psi_k}\pm i \ket{\psi_l})/\sqrt{2}$.
All terms in the right hand side of~\eqref{eq: phik phil} can be evaluated by the method described above with taking $\ket{\Phi}$ appropriately, so the $\braket{\varphi_{k}|\partial_a\partial_b\varphi_{l}}$ is also obtained.

\subsubsection{Evaluation of $\braket{\varphi_k|\partial_c\varphi_l}$}
Next, we describe how to compute $\braket{\varphi_k|\partial_c\varphi_l}$.
It follows that
\begin{equation}
\begin{split}
  \braket{\varphi_k|\partial_c\varphi_l} &= \bra{\psi_k} U^\dag(\vec{\theta}) \frac{\partial}{\partial \theta_c} U(\vec{\theta}) \ket{\psi_l} \\
  &= ig_c \bra{\psi_k} U^\dag_{1:c} P_c U_{1:c} \ket{\psi_l}.
\end{split}
\label{eq: phi del phi}
\end{equation}
The term in the last line can be evaluated by the method of Eq.~\eqref{eq: trans amp} by substituting $\ket{\varphi_{k,l}^{\pm}(\vec{R})}$ by $U_{1:c}\frac{1}{\sqrt{2}}(\ket{\psi_k}\pm\ket{\psi_l})$ and $\ket{\varphi_{k,l}^{\pm i}(\vec{R})}$ with $U_{1:c}\frac{1}{\sqrt{2}}(\ket{\psi_k}\pm i\ket{\psi_l})$.

\subsubsection{Summary}
In summary, calculation of the 2-NAC $D_{kl}^I$ proceeds as follows: 
\begin{enumerate}
 \item Perform the SSVQE or the MCVQE and obtain approximate eigenstates $\ket{\varphi_i(\vec{R})}$ and eigenenergies $\tilde{E}_i(\vec{R})$ of $H(\vec{R})$.
  \item Calculate the derivative of the Hamiltonian $\frac{\partial H}{\partial R_I}$ on classical computers and obtain $\frac{\partial\theta^*_a}{\partial R_I}$ and $\frac{\partial^2\theta^*_c}{\partial R_I^2}$ in Eq.~\eqref{eq: 2-NAC VQE} by solving Eq.~\eqref{eq: theta_first_derivative} and Eq.~\eqref{eq: theta_second_derivative}.
 \item For all $a,b=1,\ldots,N_\theta$, evaluate Eq.~\eqref{eq: PhideldelPhi} for $\ket{\Phi}=\ket{\varphi_{k,l}^\pm}, \ket{\varphi_{k,l}^{i\pm}}$,
 by using Eq.~\eqref{eq: real PhideldelPhi} and Eq.~\eqref{eq: imag PhideldelPhi}. Plugging them in Eq.~\eqref{eq: phik phil} yields the value of $\braket{\varphi_{k}|\partial_a\partial_b\varphi_{l}}$.
\item For all $c=1,\ldots,N_\theta$, evaluate $\braket{\varphi_k|\partial_c\varphi_l}$ according to Eq.~\eqref{eq: phi del phi}.
\item Substituting all values obtained in previous steps into Eq.~\eqref{eq: 2-NAC VQE} gives the 2-NAC.
\end{enumerate}
The main contribution of this paper is that we reduce the definition of the NACs (Eqs.~\eqref{eq: 1-NAC def} and \eqref{eq: 2-NAC def}) to the formulas that we can evaluate on quantum devices by the existing techniques. Here we note that the procedure 2 follows the techniques in Ref.~\cite{Mitarai2019Derivative}, the procedure 3 partially uses those in Ref.~\cite{Mitarai2019Methodology}, and the procedure 4 basically follows those in Ref.~\cite{Nakanishi2019PRR}.

\section{Calculating Berry's phase with variational quantum eigensolver \label{sec: Berry method}}
In this section, we describe a method for calculating Berry's phase with the VQE algorithm.
From the results of the VQE, while we can access the density operators of the eigenstate $\rho_k(\vec{\theta}^*)=\ket{\varphi_0(\vec{\theta}^*)}\bra{\varphi_0(\vec{\theta}^*)}$ determined by the optimized circuit-parameters $\vec{\theta}^*$, we cannot access the information about the phase of quantum state. 
Here we discuss how to calculate Berry's phase on quantum devices from the optimized circuit-parameters obtained by the VQE. 
In the following, without loss of generality, we only consider the ground state as the eigenstate.
Let $\mathcal{N}_0$ denote the set of normalized states in a complex Hilbert space $\mathcal{H}$. 
We consider performing the VQE from one point $\vec{R}_0\coloneqq \vec{R}(t_0)$ of the closed loop $\mathcal{C}_{\vec{R}}\coloneqq \{\vec{R}(t) \mid t \in [t_0, t_1], \vec{R}(t_0)=\vec{R}(t_1)\}$ in the system-parameters space and continue doing it along $\mathcal{C}_{\vec{R}}$, then we obtain a smooth curve $\mathcal{C}_{\vec{\theta}^*}\coloneqq \{\vec{\theta}^*(\vec{R}(t)) \mid t \in [t_0, t_1]\}$ in the circuit-parameter space. For simplicity, let $\vec{\theta}^*_{\mathrm{s}}\coloneqq \vec{\theta}^*(\vec{R}(t_0))$ and $\vec{\theta}^*_{\mathrm{t}}\coloneqq \vec{\theta}^*(\vec{R}(t_1))$ denote the starting point and the end point of $\mathcal{C}_{\vec{\theta}^*}$, respectively.
We note that $\vec{\theta}^*_{\mathrm{s}}\neq\vec{\theta}^*_{\mathrm{t}}$ may occur, i.e., the curve $\mathcal{C}_{\vec{\theta}^*}$ of the optimal parameters does not necessarily form the closed loop in the circuit-parameter space even when $\mathcal{C}_{\vec{R}}$ is the closed loop in the system-parameter space. 
This is because the VQE does not care about the overall phase of the ground state,
and for most cases there is a redundancy in the ansatz $\ket{\varphi_0(\vec{\theta})}$ such that $\ket{\varphi_0(\vec{\theta}_1)} = e^{i\xi} \ket{\varphi_0(\vec{\theta}_2)}, e^{i \xi} \neq 1$ for some $\vec{\theta}_1 \neq \vec{\theta}_2$.
Next, we introduce the projective Hilbert space called Ray space. 
Ray space $\mathcal{R}$ is defined as the equivalent class $\mathcal{R} \coloneqq \mathcal{N}_0/\sim$ where the equivalence relation $\sim$ holds for two elements of $\mathcal{N}_0$ which differ only by a global phase.
We also define the projection map $\pi: \ket{\psi} \in \mathcal{N}_0 \rightarrow \rho = \ket{\psi}\bra{\psi} \in \mathcal{R}$.
For a given curve $\mathcal{C}_{\mathcal{N}_0} = \{\ket{\varphi_0(\vec{\theta}^*)}\} \subset \mathcal{N}_0$, its projection to $\mathcal{R}$ is also the curve $\mathcal{C}_{\rho}\coloneqq\{\rho(\vec{\theta}^*) \mid \rho = \ket{\varphi_0(\vec{\theta}^*)}\bra{\varphi_0(\vec{\theta}^*)}, \vec{\theta}^* \in \mathcal{C}_{\vec{\theta}^*}\} \subset \mathcal{R}$, and this curve $\mathcal{C}_{\rho}$ in $\mathcal{R}$ is determined uniquely according to the optimized circuit-parameters $\vec{\theta}^*$.

Then let us describe and formulate the way to calculate Berry's phase based on the results of the VQE. Suppose that a curve $\mathcal{C}_{\rho}=\{\rho(\vec{\theta}^*)\}$ is given. Here, we consider a particular lift $\mathcal{C}_{\mathcal{N}_0}=\{\ket{\varphi_0(\vec{\theta}^*)}\}$ of $\mathcal{C}_{\rho}$ such that $\pi(\mathcal{C}_{\mathcal{N}_0})=\mathcal{C}_{\rho}$ where $\ket{\varphi_0(\vec{\theta}^*)}$ is fixed up to a phase.
We assume that $\ket{\varphi_0(\vec{\theta}^*)}$ is smooth, i.e., $\ket{\varphi_0(\vec{\theta}^*)}$ is differentiable with respect to $\vec{\theta}^*$. 
With this lift $\mathcal{C}_{\mathcal{N}_0}$, Berry's phase can be defined as \cite{MUKUNDA1993205}
\begin{equation}
\begin{split}
    \Pi_{\mathcal{C}_{\rho}} \coloneqq
    -i\int_{\vec{\theta}^*_\mathrm{s}}^{\vec{\theta}^*_\mathrm{t}}d\vec{\theta}^*\cdot&\braket{\varphi_0(\vec{\theta}^*)|\frac{\partial}{\partial\vec{\theta}^*}|\varphi_0(\vec{\theta}^*)}\\
    &\quad \quad \quad \quad \quad +\arg(\braket{\varphi_0(\vec{\theta}^*_{\mathrm{s}})|\varphi_0(\vec{\theta}^*_{\mathrm{t}})}).
\label{eq: Berry's phase based on the lift}
\end{split}
\end{equation}
We want to emphasize here that Berry's phase is a functional of the curve $\mathcal{C}_{\rho}$.
Namely, for a given curve $\mathcal{C}_{\rho}$, though we can construct a new curve $C'_{\mathcal{N}_0}$ which differs only by $U(1)$ phase degree of freedom from $\mathcal{C}_{\mathcal{N}_0}$ with a real smooth function $\Theta(\vec{\theta}^*)$,
\begin{equation}
\begin{split}
    \mathcal{C}_{\mathcal{N}_0} \rightarrow C'_{\mathcal{N}_0}: \quad & \ket{\varphi_0'(\vec{\theta}^*)}=e^{i\Theta(\vec{\theta}^*)} \ket{\varphi_0(\vec{\theta}^*)},\\
    \pi(\mathcal{C}_{\mathcal{N}_0})&=\pi(C'_{\mathcal{N}_0}),
\end{split}
\end{equation}
the value of Berry's phase $\Pi_{\mathcal{C}_{\rho}}$ calculated with the curve $\mathcal{C}'_{\mathcal{N}_0}$ is identical to that with $\mathcal{C}_{\mathcal{N}_0}$.

As discussed above, by performing the VQE, we obtain the curves $\mathcal{C}_{\vec{\theta}^*}$ and $\mathcal{C}_{\rho}$. To calculate Berry's phase based on the Eq.~\eqref{eq: Berry's phase based on the lift}, we have to choose some fixed lift $\mathcal{C}_{\mathcal{N}_0}$ from $\mathcal{C}_{\mathcal{\rho}}$. Due to the arbitrariness of the lift, we can fix the phase freedom globally and choose $\mathcal{C}_{\mathcal{N}_0}$ so that the freedom does not depend on $\vec{\theta}^*$. Therefore, given a $\rho(\vec{\theta}^*_{\mathrm{s}})$, we first choose $\ket{\varphi_0(\vec{\theta}^*_{\mathrm{s}})}$ up to phase, and then form a lift $\mathcal{C}_{\mathcal{N}_0}=\{\ket{\varphi_0(\vec{\theta}^*)}\}$ uniquely up to a phase degree of freedom in the starting point of the curve $\ket{\varphi_0(\vec{\theta}^*_\mathrm{s})}$.
By considering such lift, the terms appearing in the Eq.~\eqref{eq: Berry's phase based on the lift} can be reduced to the quantities which can be evaluated with quantum devices.
In the following, we explain how to evaluate the terms in the right hand side of Eq.~\eqref{eq: Berry's phase based on the lift}.

\subsection{Evaluation of the first term}
The first term of Eq.~\eqref{eq: Berry's phase based on the lift} is computed by discretization of the closed loop $\mathcal{C}_{\vec{R}}$ and numerical integration of the integrand.
We discretize the value of the system-parameters $\vec{R}$ on $\mathcal{C}_{\vec{R}}$ as $\vec{R}_0, \ldots, \vec{R}_{K-1}$ appropriately and also define $\vec{R}_K = \vec{R}_0$. 
The VQE algorithm is performed for all points $\{\vec{R}_p\}_{p=0}^K$ and the optimal circuit-parameters are obtained as $\{\vec{\theta}^*_p = \vec{\theta}^*(\vec{R}_p) \}_{p=0}^K$.
We define $\vec{\theta}_0^* \coloneqq \vec{\theta}^*_\mr{s}(\vec{R}_0)$ and $\vec{\theta}_K^* \coloneqq \vec{\theta}^*_\mr{t}(\vec{R}_0)$
and stress again that $\vec{\theta}_0^* \neq \vec{\theta}_K^*$ may hold in general due to the redundancy of the ansatz.
Here because we choose the phase freedom of the lift $\mathcal{C}_{\mathcal{N}_0}=\{\ket{\varphi_0(\vec{\theta}^*)}\}$ which is independent on $\vec{\theta}^*$, the integrand of the first term can be written as
\begin{equation}
\begin{split}
\braket{\varphi_0(\vec{\theta}^*)|\frac{\partial}{\partial \theta_a^*}|\varphi_0(\vec{\theta}^*)}
&= \bra{\psi_0} U^\dag(\vec{\theta}^*) \frac{\partial}{\partial \theta_a^*} U(\vec{\theta}^*) \ket{\psi_0} \\
&= ig_a \bra{\psi_0} U^{*\dag}_{1:a} P_a U_{1:a}^* \ket{\psi_0},
\end{split}
\label{eq: Berry integrand term}
\end{equation}so it is evaluated by measuring the expectation value of $P_a$ for the state $U_{1:a}^*\ket{\psi_0}$, which can be evaluated on quantum devices.
Therefore the integral is approximated by
\begin{eqnarray}
 && \int_\mathcal{C'} d\vec{\theta}^* \cdot
 \braket{\varphi_0(\vec{\theta}^*)|\frac{\partial}{\partial\vec{\theta}^*}|\varphi_0(\vec{\theta}^*)}\nonumber \\
&\approx&
 \sum_{p=0}^{K-1} (\vec{\theta}_{p+1}^* -\vec{\theta}_p^*) \cdot
 \left. \braket{\varphi_0(\vec{\theta}^*)|\frac{\partial}{\partial\vec{\theta}^*}|\varphi_0(\vec{\theta}^*)}\right|_{\vec{\theta}^*=\vec{\theta}^*_p}.
\label{eq: Berry theta sum}
\end{eqnarray}

\subsection{Evaluation of the second term}
\begin{figure}
\includegraphics[width=0.44\textwidth]{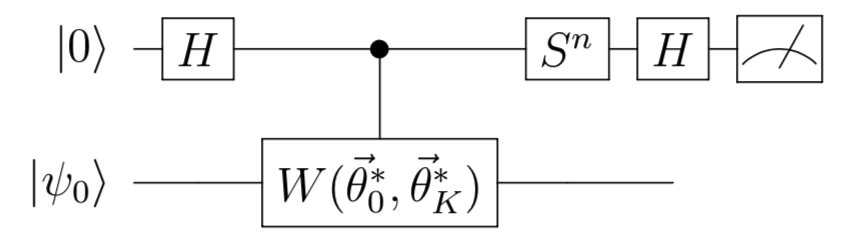}
\caption{The Hadamard test to evaluate the phase difference $\arg(\braket{\varphi(\vec{\theta}^*_{\mathrm{s}})|\varphi(\vec{\theta}^*_{\mathrm{t}})})$ in Eq.~\eqref{eq: Berry's phase based on the lift}. The upper line represents an ancillary qubit which is measured, and the lower line represents the system on which $W(\vec{\theta}_0^*,\vec{\theta}_K^*)\coloneqq U^{\dag}(\vec{\theta}^*_K)U(\vec{\theta}^*_0)$ operates.
The results of the measurements for the ancillary qubit gives the value of  $\mr{Re}\left(\braket{\varphi(\vec{\theta}_0^*)|\varphi(\vec{\theta}_K^*)}\right)$ and $\mr{Im}\left(\braket{\varphi(\vec{\theta}_0^*)|\varphi(\vec{\theta}_K^*)}\right)$ for $n=0, 1$.}
\label{fig: Hadamard test for berry}
\end{figure}

The second term in Eq.~\eqref{eq: Berry's phase based on the lift}, $\arg(\braket{\varphi_0(\vec{\theta}^*_{\mathrm{s}})|\varphi_0(\vec{\theta}^*_{\mathrm{t}})})$, is evaluated by the difference of the overall phase of two wavefunctions $\ket{\varphi_0(\theta^*_0)}$ and $\ket{\varphi_0(\theta^*_K)}$.
This can be performed by estimating $\arg(\braket{\varphi_0(\theta^*_0)|\varphi_0(\theta^*_K)}) = \arg(\braket{\psi_0|U^{\dag}(\vec{\theta}^*_0)U(\vec{\theta}^*_K)|\psi_0})$ with the Hadamard test~\cite{1998RSPSA.454..339C} depicted in Fig.~\ref{fig: Hadamard test for berry}.
It requires one ancillary qubit and the controlled-$U(\vec{\theta})$ gates, which are costly for near-term quantum devices.
We finally mention that if we construct the lift $\mathcal{C}_{\mathcal{N}_0}$ of $\mathcal{C}_{\rho}$ so that the phase degree of freedom depends on $\vec{\theta}^*$, by taking both terms in Eq.~\eqref{eq: Berry's phase based on the lift} into account, Eq.~\eqref{eq: Berry's phase based on the lift} can be reduced to the quantities which can be evaluated on quantum devices as discussed above~\footnote{
Even if we choose the phase degree of freedom that depends on $\vec{\theta}^*$, the quantities that are evaluated on quantum devices to calculate Berry's phase is the same. We consider the lift $\mathcal{C}'_{\mathcal{N}_0}=\{e^{-i\Theta(\vec{\theta}^*)}\ket{\varphi_0(\vec{\theta}^*)}\}$ with $\Theta_\mathrm{s(t)}\coloneqq \Theta(\vec{\theta}^*_{\mathrm{s(t)}})$, and the integrand of the first term can be written as $(\bra{\varphi_0(\vec{\theta}^*)}e^{i\Theta(\vec{\theta}^*)})\frac{\partial}{\partial \theta_a^*}(e^{-i\Theta(\vec{\theta}^*)}\ket{\varphi_0(\vec{\theta}^*)})=-ie^{i\Theta(\vec{\theta}^*)}\cdot\frac{\partial \Theta(\vec{\theta}^*)}{\partial \theta_a}+\bra{\psi_0} U^\dag(\vec{\theta}^*) \frac{\partial}{\partial \theta_a^*} U(\vec{\theta}^*) \ket{\psi_0}$. Integrating this term over $\mathcal{C}_{\vec{\theta}^*}$ gives additional terms $(\Theta_{\mathrm{t}}-\Theta_{\mathrm{s}})$ with respect to the first term of Eq.~\eqref{eq: Berry's phase based on the lift}, but, on the other hand, $\arg({\bra{\varphi_0(\vec{\theta}^*_\mathrm{s})}}e^{i\Theta_{\mathrm{s}}}\cdot e^{-i\Theta_{\mathrm{t}}}\ket{\varphi_0(\vec{\theta}^*_\mathrm{t})})$ also gives rise to additional terms $-(\Theta_{\mathrm{t}}-\Theta_{\mathrm{s}})$, which cancel each other out.}.

\subsection{Comparison with previous studies}
We here compare previous work on calculating Berry's phase on quantum devices with our method.
In Ref.~\cite{Murta2020}, Berry's phase is calculated by simulating adiabatic dynamics of the system $U_\mathcal{C} = \mathcal{T}e^{-i\int_0^T ds H(s)}$, where $\mathcal{T}$ is the time-ordered product and $H(s)$ is a time-dependent Hamiltonian, which varies along the closed loop $\mathcal{C}$ in sufficiently long time $T$.
$U_\mathcal{C}$ is implemented on quantum computers by the Suzuki-Trotter decomposition, and the Hadamard test like Fig.~\ref{fig: Hadamard test for berry} is performed to detect the phase difference between the initial ground state $\ket{\chi(\vec{R}_0)}$ and the time-evolved state $U_\mathcal{C}\ket{\chi(\vec{R_0})}$.
The phase difference between $\ket{\chi(\vec{R}_0)}$ and $U_\mathcal{C}\ket{\chi(\vec{R_0})}$ contains the dynamical phase and Berry's phase, but the former phase can be neglected by combining the forward- and backward-time evolutions by assuming the Hamiltonian of the system has the time-reversal symmetry.
Compared with this strategy, our proposal for calculating Berry's phase based on the VQE has two features.
First, we do not have to assume the time-reversal symmetry in the system to remove the contribution from the dynamical phase like in the previous study because we directly calculate Berry's phase based on the definition Eq.~\eqref{eq: def Berry}.
Our method can apply to general quantum systems.
Second, the causes of errors are quite different. More concretely, while the errors in the previous methods arise from the Trotterization of the time evolution operator, the errors in our method mainly come from two sources: one is the approximation error of the eigenstates obtained by the VQE, and the other is the numerical error of integration in Eq.~\eqref{eq: Berry theta sum}. These errors can be reduced by deepening the ansatz circuits and taking more discretized points on $\mathcal{C}$, respectively.
We comment that further research is needed to conclude the difference in the performance between our method and these previous methods.

Finally, we introduce another method to calculate Berry's phase based on the VQE with a lot of Hadamard tests.
Using the formula
\begin{equation}
 \Pi_\mathcal{C} \approx - i \sum_{i=0}^K \mr{Im} \left( \ln  \braket{\varphi_0(\vec{\theta}^*_i)|\varphi_0(\vec{\theta}^*_{i+1})} \right),
\label{eq: Fukui-Hatsugai}
\end{equation}
with taking the principal branch of the complex logarithm, $-\pi \leq \mr{Im}(z) < \pi$ for $z\in\mathbb{C}$, is one of the candidates for avoiding discretization error of the closed loop $\mathcal{C}$ and numerical instability~\cite{Fukui2005}.
The value of $ \braket{\varphi_0(\vec{\theta}^*_i)|\varphi_0(\vec{\theta}^*_{i+1})}$ is evaluated by the Hadamard test in Fig.~\ref{fig: Hadamard test for berry} by substituting $\vec{\theta}_{0(K)}^*$ with $\vec{\theta}_{i(i+1)}^*$.

\subsection{Summary}
Berry's phase can be calculated based on the VQE as follows:
\begin{enumerate}
 \item Discretize the closed loop $\mathcal{C}$ in the system-parameters space as $\{\vec{R}_p\}_{p=0}^K$ appropriately and perform the VQE for all points.
 \item Calculate the first term of Eq.~\eqref{eq: Berry's phase based on the lift} by using Eq.~\eqref{eq: Berry integrand term} and Eq.~\eqref{eq: Berry theta sum}.
 \item If necessary, evaluate the phase difference $\arg(\braket{\varphi_0(\vec{\theta}^*_{\mathrm{s}})|\varphi_0(\vec{\theta}^*_{\mathrm{t}})})$ by the Hadamard test shown in Fig.~\ref{fig: Hadamard test for berry}.
 \item Substituting all values obtained in the previous steps into Eq.~\eqref{eq: Berry's phase based on the lift} gives the Berry's phase.
\end{enumerate}

\section{Experiment on a real quantum device \label{sec: experiment}}
\begin{figure*}
\includegraphics[width=1.\textwidth]{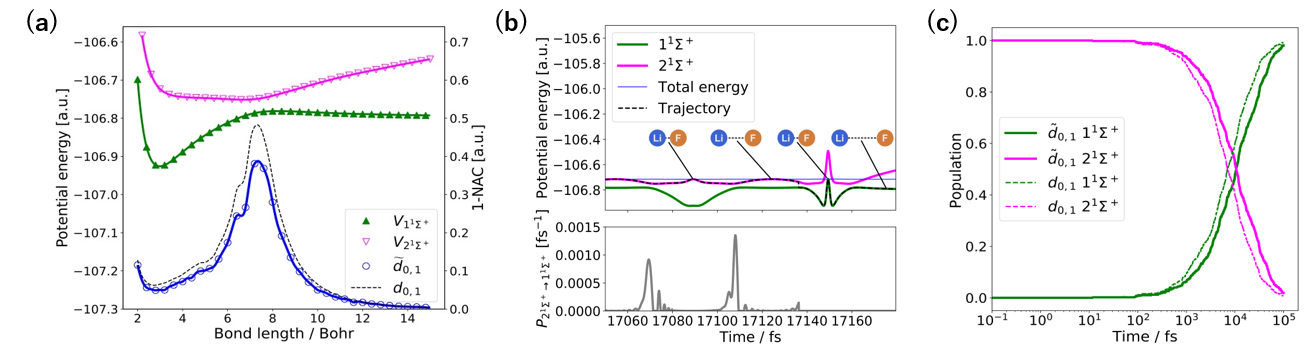}
\caption{
(a) Numerical results of the potential energy curves of the $1^1\Sigma^+$ state (upper triangles) and $2^1\Sigma^+$ state (lower triangles) of LiF obtained by the SSVQE and experimental results of the 1-NAC $\tilde{d}_{0,1}$ between the two states (circles) at the bond distance $R$ in the range of $2$ to $15$ Bohr. The solid lines represent the spline interpolated curves computed by SciPy, a numerical library in Python~\cite{2020NatMe..17..261V}. The dashed lines show the exact numerical results of the 1-NAC $d_{0,1}$ in a noiseless situation calculated with the Qiskit's statevector simulator~\cite{Qiskit2020}.
The exact potential energy curves calculated by the exact diagonalization are not presented in the figure since they coincide with the ones obtained by the SSVQE (presented in the figure) by less than $10^{-10}$ a.u. for the whole range of $R$. 
(b) Potential energies of $1 ^1\Sigma^+$ state and $2 ^1\Sigma^+$ state (upper panel) of LiF and the transition probability $P_{2^1\Sigma^+ \rightarrow 1^1\Sigma^+}$ (lower panel) for one trajectory obtained by the TSH molecular dynamics simulation using $\tilde{d}_{0,1}$ values in (a). On this trajectory, the molecule hops from $2 ^1\Sigma^+$ state to $1 ^1\Sigma^+$ state at 17136.2 fs during bond shrinking motion and it is dissociated into Li and F atoms. 
(c) Time evolution of the population of the $1 ^1\Sigma^+$ state and $2 ^1\Sigma^+$ state of LiF computed by 500 trajectories of TSH simulation using the values of 1-NAC, $\tilde{d}_{0,1}$ and $d_{0,1}$, in (a).
\label{fig: exp NAC}}
\end{figure*}

In this section, we show an experimental result of our algorithm for the 1-NAC on a real quantum device and the non-adiabatic molecular dynamics (MD) simulation based on the experimental values of the 1-NAC.

We consider a lithium fluoride (LiF) molecule with bond length $R$ and its electronic states under the Born-Oppenheimer approximation.
In this system, the potential energy curves, or eigenenergies $E$ as a function of $R$, of the two lowest $^1\Sigma^+$ states are known to exhibit the avoided crossing \cite{Werner1981, Bauschlicher1988, Bandrauk1989, Bandrauk1990}, and it plays a crucial role for nonadiabatic dynamics such as photodissociation.
Here we focus on this avoided crossing and the resulting nonadiabaitc dynamics by modeling the system with a simple two-state model.
Specifically, the electronic Hamiltonian of LiF at bond distance $R$ is constructed by two orbitals obtained by the state-average complete active space self-consistent field (CASSCF) method \cite{Sun2017}.
By considering symmetries in the system, one can obtain a two-qubit Hamiltonian $H_{\mr{LiF}}(R)$ from that electronic Hamiltonian.
Further details are described in Appendix~\ref{app: Ham LiF}.

\begin{figure}
\includegraphics[width=.45\textwidth]{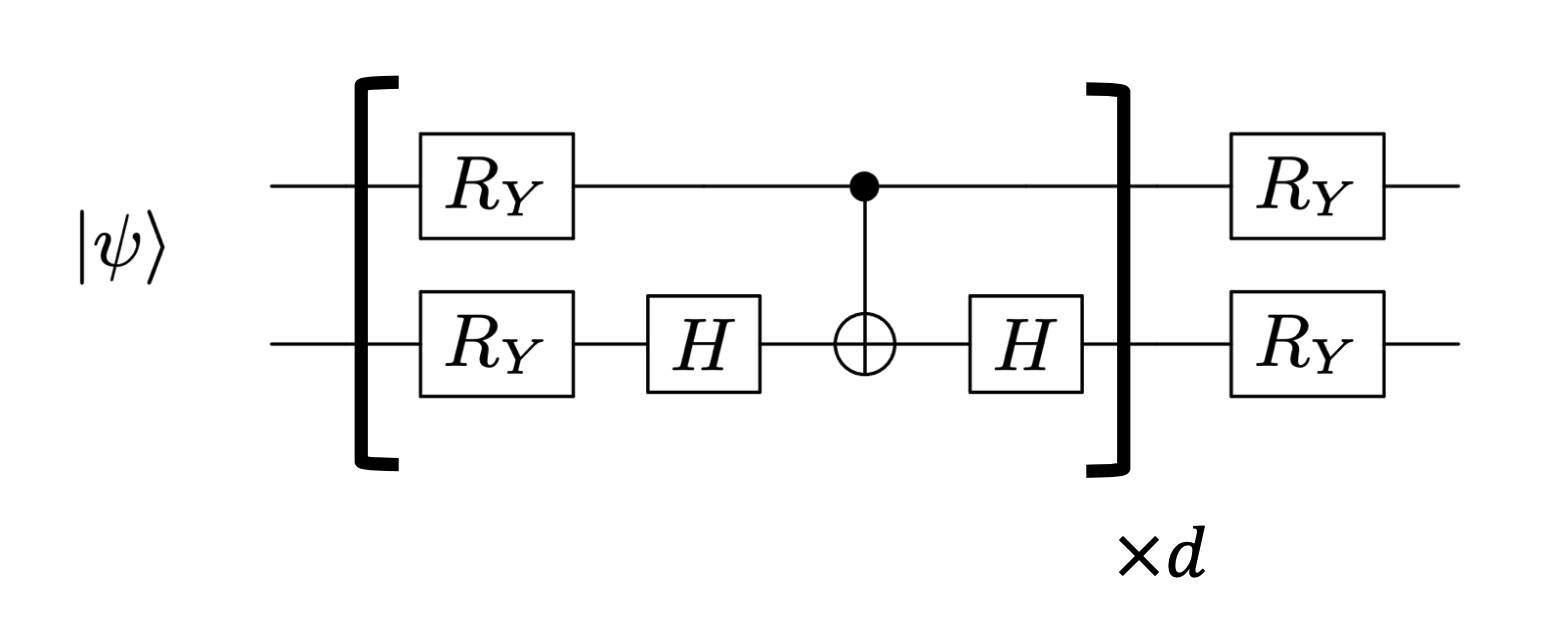}
\caption{
The ansatz circuit $U(\vec{\theta})$ for the SSVQE.
Each $R_Y = \exp(-iY\theta/2)
= \begin{pmatrix}
\cos\theta/2 & \sin\theta/2 \\
-\sin\theta/2 & \cos\theta/2 \\
\end{pmatrix}$ has a rotational angle $\theta$ as an independent parameter, and $d=4$ denotes the depth of the ansatz.
The rotation angles of $R_Y$ gates are optimized during the SSVQE in our experiments.
We note that quantum states generated by this circuit for real wavefunction $\ket{\psi}$ remain real for any choice of $\theta$.
\label{fig: ansatz}}
\end{figure}

We run our algorithm to calculate the 1-NAC (Sec.~\ref{sec: NAC method}) of $H_{\mr{LiF}}(R)$ at various bond length $R$ in the IBM Q cloud quantum device (\verb|ibmq_valencia|)~\cite{IBMQ2020}.
First, the SSVQE calculation for $H_{\mr{LiF}}(R)$ is performed on a classical simulator where the exact and noiseless expectation values of observables are obtained.
The ansatz for the SSVQE, $U(\vec{\theta})$, is depicted in Fig.~\ref{fig: ansatz}
and we optimize the parameters $\vec{\theta}$ to minimize the cost function~\eqref{eq: cost of SSVQE}.
The initial states to which the ansatz applied are $\ket{\psi_0}=\ket{00}, \ket{\psi_1} = \ket{01}$ states. 
To calculate the singlet ($S=0$) states $^1\Sigma^+$, we penalize the triplet ($S=1$) states by modifying the Hamiltonian in the cost function as $H' = H_{\mr{LiF}}(R) + \beta \hat{S}^2$, where $\hat{S}^2$ is the spin squared operator and $\beta=4.0$ is a constant~\cite{kuroiwa2021}. 
We obtain the (approximate) eigenstates $\ket{\varphi_{0,1}(\vec{\theta}^*)}=U(\vec{\theta}^*)\ket{\psi_{0,1}}$ and eigenenergies $\tilde{E}_{0,1}$ for two $^1\Sigma^+$ states in this way from the classical simulator~(Fig.~\ref{fig: exp NAC}(a) shows the potential energy curves).
Next, the transition amplitude $\braket{\varphi_0(\vec{\theta}^*)|\frac{dH_{\mr{LiF}}(R)}{dR}|\varphi_1(\vec{\theta}^*)}$ is computed on a quantum device by evaluating each term of the right hand sides of Eq.~\eqref{eq: trans amp} with 8192 shots.
Since the wavefunction is real by definition of the ansatz $U(\vec{\theta})$, we consider only the real part of Eq.~\eqref{eq: trans amp}.
Finally, the 1-NAC is obtained by plugging the estimate of  $\braket{\varphi_0(\vec{\theta}^*)|\frac{dH_{\mr{LiF}}(R)}{dR}|\varphi_1(\vec{\theta}^*)}$ into the numerator of Eq.~\eqref{eq: 1-NAC formula} and that of $\tilde{E}_0 -\tilde{E}_1$ into the denominator.
As we mentioned in Sec.~\ref{sec: review of NDC}, the 1-NAC is not gauge invariant.
Even when considering only real wavefunctions as in this case, the 1-NAC still has indefinite sign, but here the sign is determined by considering the continuity of the 1-NAC with respect to the nuclear coordinate $\vec{R}$.
We comment that the way of calculating the 1-NAC here is chosen to remove the effect of the noise and errors in the real quantum device during the SSVQE and focus on evaluating the transition amplitude.

The result of the 1-NAC is shown in Fig.~\ref{fig: exp NAC}(a).
Because of the noise in the real quantum device, the values of the transition amplitude are smaller than the exact results but still qualitatively consistent with them.
This shrinking could be resolved by, for example, using the error mitigation technique~\cite{temme2017error,endo2018practical} for near-term quantum devices.

In addition, we perform the trajectory surface hopping (TSH)~\cite{Tully1971} molecular dynamics calculation using Tully’s fewest switches algorithm~\cite{Tully1990} based on the obtained values of the potential energy curves and the 1-NAC (Fig.~\ref{fig: exp NAC}(a)).
We assume a situation where a LiF molecule is excited by light to the first electronic excited state. 
In the TSH simulation, the nuclear energy gradient $\frac{\partial \tilde{E}_{0,1}}{\partial R}$ and the 1-NAC at various bond lengths $R$ are requested by the TSH program code, and we feed it with the values interpolated from the results of the SSVQE and the 1-NAC experiment.
The details of the interpolation are described in Appendix~\ref{app: interpolation}.
We prepare a set of 500 molecular geometries and nuclear velocities as a harmonic-oscillator Wigner distribution for the vibrational ground state at the equilibrium geometrical structure in the electronic ground state $1^1\Sigma^+$.
We run the trajectories from the first electronic excited state $2^1\Sigma^+$ with a time step of 0.1 fs and find that the trajectories hop from $2^1\Sigma^+$ to $1^1\Sigma^+$ where the dissociation occurs as shown in Fig.~\ref{fig: exp NAC}(b).
The dynamics of the populations of the $1^1\Sigma^+$ state and $2^1\Sigma^+$ state is calculated based on the results of 500 trajectories and shown in Fig.~\ref{fig: exp NAC}(c).
Since the experimentally-obtained 1-NAC values are smaller than the noiseless simulation ones (Fig.~\ref{fig: exp NAC}(a)), the decay of the $2^1\Sigma^+$ population calculated by using the  experimental values of 1-NAC is slightly slower than that is calculated by using the noiseless simulation values of 1-NAC.
Nevertheless, the overall dynamics of the populations is similar to each other and 
this indicates the possibility of performing TSH in a quantum device in the near future.

The TSH simulation is conducted by the open-source library SHARC~\cite{sharc-md2.1, Richter2011JCTC, https://doi.org/10.1002/wcms.1370}.

Although we consider the 1-NAC throughout this section, we perform additional numerical demonstrations of our methods by simulating the quantum circuits to calculate the 1-NAC and 2-NAC of the hydrogen molecules and Berry's phase of a two-site spin model in Appendix~\ref{app: numerical simulation}.

\section{Discussion \label{sec: disussion}}
Our proposed methods presented in the previous sections are based on the analytical derivative of the eigenstates obtained by the SSVQE.
This section compares our methods with those using the numerical derivative of the eigenstates by the finite difference method.
The comparison will be made from two points of view:
(1) the number of distinct Hamiltonians to perform the SSVQE to obtain the optimal circuit-parameters $\vec{\theta}^{*}$ and (2) the total number of measurements required to evaluate the NACs and Berry's phase after performing the SSVQE.
We ignore the cost of classical computation throughout analyses in this section.

We recall that $N_\theta$ and $N_x$ are the dimensions of the circuit-parameters and system-parameters, respectively.
The number of qubits in the system is denoted as $n$ and the number of Pauli terms in the Hamiltonian as $N_H$.
For quantum chemistry problems $N_H$ is typically $O(n^4)$~\cite{RevModPhys.92.015003, Cao2019}, but several methods for reducing $N_H$ are proposed~\cite{Motta2018npj, 2021npjQI...7...23H}.
The detailed derivations of the formulas in this section (Eqs.~\eqref{eq: main 1-NAC cost},~\eqref{eq: main 1-NAC cost finite}, and \eqref{eq: main cost 2-NAC finite}) are presented in Appendix~\ref{app: cost ours}.

\subsection{Cost of 1-NAC}
Let us consider our proposed method to calculate the 1-NAC $d_{kl}^I$ with fixed $k,l$ and all $I=1,\ldots N_x$ at some fixed system-parameters $\vec{R}$.
The evaluation of the 1-NAC is performed by calculating the denominator and numerator in Eq.~\eqref{eq: 1-NAC formula}.
Both can be obtained by applying a single optimized ansatz circuit $U(\vec{\theta}^*(\vec{R}))$ resulting from the SSVQE to several initial states and measuring appropriate observables, so the number of distinct Hamiltonians to perform the SSVQE is just one.
Meanwhile, the number of measurements to calculate the 1-NAC is estimated as follows.
When we write the Hamiltonian as $H(\vec{R}) = \sum_{i=1}^{N_H} h_i(\vec{R}) P_i$, where $P_i$ is Pauli operator and $h_i(\vec{R})$ is a real coefficient, the values of $\braket{\varphi_{k(l)}|H(\vec{R})|\varphi_{k(l)}} = \sum_{i=1}^{N_H} h_i(\vec{R}) \braket{\varphi_{k(l)}| P_i|\varphi_{k(l)}}$ and $\braket{\varphi_k|\partial H(\vec{R})/\partial R_I|\varphi_l} = \sum_{i=1}^{N_H} \partial h_i(\vec{R})/\partial R_I \braket{\varphi_k|P_i|\varphi_l}$ are necessary to compute the 1-NAC.
The term $\braket{\varphi_{k(l)}|P_i|\varphi_{k(l)}}$ is evaluated as the expectation value of $P_i$, and similarly the term $\braket{\varphi_k|P_i|\varphi_l}$ ($i=1,\ldots,N_H$) is evaluated as a sum of four expectation values in the right hand sides of Eq.~\eqref{eq: trans amp}.
By taking into account errors in evaluating those expectations values, the number of measurements to estimate the 1-NAC within the error $\epsilon$ is given by 
\begin{equation}
\begin{split}
 &N_{\mr{total}}^\mr{1\mathchar`-NAC}  \\
 &= O\left[\frac{N_H}{\epsilon^2 |\Delta E_{k,l}|^4 }\left( |\Delta E_{k,l}| \left\| \frac{\partial H}{\partial R_{I^*}}\right\| + \|H\| |A_{I^*}| \right)^2 \right],
\end{split}
\label{eq: main 1-NAC cost}
\end{equation}
where $A_I (\Delta E_{k,l})$ is the numerator (denominator) of Eq.~\eqref{eq: 1-NAC formula}, $\|H\| = \sum_i |h_i|$, $\| \partial H /\partial R_I \| = \sum_i |\partial h_i /\partial R_I|$, and $I^* = \mr{argmax}_I \left( |\Delta E_{k,l}| \left\| \frac{\partial H}{\partial R_{I}}\right\| + \|H\| |A_{I}| \right)$.
It scales with the number of the Pauli terms in the Hamiltonian but does not depend on the number of circuit-parameters $N_\theta$ by virtue of Eq.~\eqref{eq: 1-NAC formula}.
The dependence on $I$, or the system-parameters, is also absent because it is absorbed into the classical computation of the coefficient $\partial h_i(\vec{R})/\partial R_I$~\cite{Mitarai2019Derivative}.

To compare with our method, one can consider a method to evaluate the 1-NAC based on numerical differentiation of the eigenstates obtained by the SSVQE.
In such approach, the 1-NAC can be evaluated by the following formula, 
\begin{equation}
    d_{k,l}^{I} \approx \frac{\tau_{k,l}(\vec{R}, \vec{R}+ h\vec{e}_I)- \tau_{k,l}(\vec{R}, \vec{R}- h\vec{e}_I)}{2h},
    \label{eq: numerical differentiation of 1-NAC}
\end{equation}
where $\tau_{k,l}(\vec{R}, \vec{R}\pm h\vec{e}_I)$=$\braket{\varphi_k(\vec{\theta}^*(\vec{R}))|\varphi_l(\vec{\theta}^*(\vec{R}\pm h\vec{e}_I))}$, $h$ is a positive number, and $\vec{e}_I$ is the unit vector in $I$-th direction. For simplicity, we will represent $\tau_{k,l}(\vec{R}, \vec{R}\pm h\vec{e}_I)$ as $\tau_{k,l}^{\pm, I}$.

When evaluating each term of Eq.~\eqref{eq: numerical differentiation of 1-NAC}, we assume that ${\tau_{k,l}^{\pm,I}}$ is estimated from the overlap $\left|\braket{\psi_k|U^{\dag}(\vec{\theta}^{*}(\vec{R}))U(\vec{\theta}^{*}(\vec{R}\pm h\vec{e}_I))|\psi_l}\right|^2$, which can be easily evaluated from measurements on near-term quantum devices if $\ket{\psi_k}, \ket{\psi_l}$ are computational basis states~\cite{Higgott2019}.
We then obtain the value of $\tau_{k,l}^{\pm, I}$ by taking the square root of the overlaps with a positive sign (real value)~\cite{Havlicek2019}.
This treatment can be justified when we solve problems in quantum chemistry, where wavefunctions are often real, and we adopt an ansatz which produces a real wavefunction.
This approach to evaluate Eq.~\eqref{eq: numerical differentiation of 1-NAC} avoids the costly Hadamard test and is considered to be feasible on near-term quantum devices.
We note that our methods are always applicable without the assumption above.

To evaluate the 1-NAC with Eq.~\eqref{eq: numerical differentiation of 1-NAC}, we need optimal parameters $\vec{\theta}^{*}(\vec{R}), \vec{\theta}^{*}(\vec{R}\pm h\vec{e}_I)$ for all $I$, so the number of distinct Hamiltonians to perform the SSVQE in the finite difference method is $2N_x + 1 = O(N_x)$.
By considering the error in estimating the overlaps, the number of measurements to estimate the 1-NAC with the precision of $\epsilon$ in the finite difference method is at least
\begin{equation}
    {N^{'}}_{\mr{total}}^\mr{1\mathchar`-NAC}
    = O(N_x/T_{k,l}^2\epsilon^2), \label{eq: main 1-NAC cost finite}
\end{equation}
where $T_{k,l}= \min_{\sigma=\pm, I} \tau_{k,l}^{\sigma,I}$ and we assume the condition $M_3 \geq O(\epsilon)$, where
$M_3 = \max_{I} \max_{s\in[-h,h]} \left|\tau_{k,l}^{(I, 3)}(s) \right|$ and $\tau_{k,l}^{(I, 3)}(s) = \frac{d^3}{ds^3} \tau_{k,l}(\vec{R}, \vec{R}+s\vec{e}_I)$.
Both our method~\eqref{eq: main 1-NAC cost} and the finite difference method~\eqref{eq: main 1-NAC cost finite} scale with $\epsilon^{-2}$, so the prefactors determine the efficiency of them.
When $N_x$, or the number of system-parameters (nuclei of the molecule), becomes large, the finite difference method will suffer from a large number of the SSVQE runs and the measurements compared with our method.  

\subsection{Cost of 2-NAC}
To calculate the 2-NAC $D_{kl}^I$ with fixed $k,l$ and all $I=1,\ldots N_x$ at $\vec{R}$ with our method, we require one optimized circuit-parameter $\theta^{*}(\vec{R})$, so the number of distinct Hamiltonians to perform the SSVQE is again one.
Let us consider the number of measurements.
We need the derivatives $\{ \frac{\partial\theta^*_a}{\partial R_I}, \frac{\partial^2\theta^*_a}{\partial R_I^2}\}_{I=1}^{N_x}$ which are obtained as the solutions of Eqs.~\eqref{eq: theta_first_derivative}\eqref{eq: theta_second_derivative}.
The coefficients of these equations are determined within error $\epsilon$ by performing $O(N_\theta^3 N_H/\epsilon^2)$ measurements.
Since the error propagation from the coefficients to the solutions $\{ \frac{\partial\theta^*_a}{\partial R_I}, \frac{\partial^2\theta^*_a}{\partial R_I^2}\}_{I=1}^{N_x}$ is very complicated, we here let the error of the solutions be $\epsilon$ (see Ref.~\cite{Mitarai2019Derivative} for similar discussion).
The value of $\braket{\varphi_{k}|\partial_a\partial_b\varphi_{l}} (a,b=1,\ldots,N_\theta)$ in Eq.~\eqref{eq: phik phil} is obtained by measuring $O(N_\theta^2/\epsilon^2)$ times within error $\epsilon$.
Similarly, $\braket{\varphi_{k}|\partial_c\varphi_{l}} (c=1,\ldots,N_\theta)$ in Eq.~\eqref{eq: phi del phi} is calculated by $O(N_\theta/\epsilon^2)$ measurements.
Therefore, the total number of measurements to evaluate the 2-NAC by our method is roughly given by
\begin{equation}
N_{\mr{total}}^{\mr{2\mathchar`-NAC}}=O(N_\theta^3 N_H/\epsilon^2), \label{eq: main cost 2-NAC}
\end{equation}
where the meaning of $\epsilon$ has to be cared for.
It does not depend on the number of system-parameters $N_x$.

The finite difference method is based on the following formula:
\begin{equation}
    D_{k,l}^{I} \approx \frac{\tau_{k,l}(\vec{R}, \vec{R}+ h\vec{e}_I)+ \tau_{k,l}(\vec{R}, \vec{R}- h\vec{e}_I)}{h^2}
    \label{eq: numerical differentiation of 2-NAC}
\end{equation}
where we used $\tau_{k,l}(\vec{R}, \vec{R})=0$.
Similarly to the case of the 1-NAC, the number of distinct Hamiltonians to perform the SSVQE is $O(N_x)$.
To bound the error of the 2-NAC by $\epsilon$, the finite difference method requires at least
\begin{equation}
 {N^{'}}_{\mr{total}}^{\mr{2\mathchar`-NAC}}=O\left(N_x^2/T_{k,l}^2 \epsilon^2 \right)
 \label{eq: main cost 2-NAC finite}
\end{equation}
measurements under the condition $M_4\geq O(\epsilon)$, where
$M_4 = \max_I \max_{s\in[-h,h]} \left|\tau_{k,l}^{(I, 4)}(s) \right|$ and $\tau_{k,l}^{(I, 4)}(s) = \frac{d^4}{ds^4} \tau_{k,l}(\vec{R}, \vec{R}+s\vec{e}_I)$.
The number of measurements in our method~\eqref{eq: main cost 2-NAC} does not depends on $N_x$ while the finite difference method~\eqref{eq: main cost 2-NAC finite} does, as with the 1-NAC. 

\subsection{Cost of Berry's phase}
For calculating  Berry's phase, the closed path $\mathcal{C}$ is discretized into $K$ points. The integrand (Eq.~\eqref{eq: Berry's phase based on the lift}) is evaluated at all $K$ points and numerically integrated both in our method and the finite difference method.
We therefore compare our method with the finite difference method only in terms of the cost to obtain the integrand at all the discretized points.

In our method, the integrand at each discretized point can be evaluated by Eq.~\eqref{eq: Berry integrand term}.
The total number of distinct Hamiltonians to perform the VQE is $K+1 = O(K)$.
If we bound the error in estimating each integrand by $\epsilon$, the total number of measurements is given by
\begin{equation}
    N_{\mr{total}}^{\mr{Berry}}=O(KN_\theta/\epsilon^2).
\end{equation}

In the finite difference, the integrand can be evaluated with the finite difference method by the following formula,
\begin{equation}
\begin{split}
\frac{\braket{\varphi_0(\vec{R}_k)|\varphi_0(\vec{R}_k+ h\vec{v}_k)}-\braket{\varphi_0(\vec{R}_k)|\varphi_0(\vec{R}_k- h\vec{v}_k)}}{2h}
\end{split}
\end{equation}
where $\vec{v}_k \propto \vec{R}_{k+1} - \vec{R}_k$ is the unit vector along the closed loop $\mathcal{C}$.
The minimal number of measurements to obtain all integrands within error $\epsilon$ can be derived in a similar way for the 1-NAC, and the result is 
\begin{equation}
    {N^{'}}_{\mr{total}}^{\mr{Berry}}=O(K/T_{0,0}^2\epsilon^2),
\end{equation}
under the condition $M_3^{'}\geq O(\epsilon)$ where $M_3^{'} = \max_k \max_{s\in[-h,h]} \left|\tau_{0,0}^{(k, 3)}(s) \right|$ and $\tau_{0,0}^{(k, 3)}(s) = \frac{d^3}{ds^3} \tau_{0,0}(\vec{R}, \vec{R}+s\vec{v}_k)$.

\section{Conclusion \label{sec: conclusion}}
In this paper, we have proposed methods to calculate the NACs and Berry's phase based on the VQE.
We utilize the SSVQE and the MCVQE algorithms, which enable us to evaluate transition amplitudes of observables between approximate eigenstates.
We explicitly present quantum circuits and classical post-processings to evaluate the NACs and Berry's phase in the framework of the VQE.
For the 1-NAC, the calculations are simplified by taking advantage of the formula~\eqref{eq: 1-NAC formula}.
The 2-NAC is obtained by combining the projective measurements and the expectation-value measurements of Pauli operators.
The evaluation of Berry's phase is also carried out by the measurements of expectation values of Pauli operators with numerical integration of the definition of Berry's phase in addition to performing the Hadamard test once.
We note that our method for calculating Berry's phase is applicable for molecular systems which have the conical intersection~\cite{doi:https://doi.org/10.1002/0471780081.ch4}.
To show the potential feasibility of our method for the 1-NAC on a near-term quantum device, we evaluate the value of the 1-NAC of a lithium fluoride molecule on the IBM Q processor. Based on those results, we perform the nonadiabatic molecular dynamics simulation of photodissociation of a lithium fluoride for the first time.
The methods given in the present paper contribute to enlarging the usage of the VQE and accelerate further developments to investigate quantum chemistry and quantum many-body problems on near-term quantum devices.

We lastly comment on the effect of the barren plateau problem~\cite{2018NatCo...9.4812M, 2021NatCo..12.1791C, Cerezo_2021, Uvarov_2021, sharma2020trainability} to our methods.
The barren plateau problem states that the gradients of the expectation values of observables with respect to ansatz circuit parameters vanish exponentially with the increase of the number of qubits when the ansatz has enough expressibility.
When the barren plateau occurs, it is difficult to obtain the optimal circuit parameters $\vec{\theta}^*$ because the gradients become too small to optimize the parameters.
Our proposed methods in this paper totally discuss the procedures after the optimal circuit parameters have been obtained (i.e., the VQE has successfully converged).
Although our methods do not work when we cannot obtain $\vec{\theta}^*$, several techniques~\cite{Grant_2019, Skolik_article, verdon2019learning, anand2020natural, pesah2020absence} to avoid or ameliorate the barren plateau for the VQE and other variational quantum algorithms have been proposed.

\section*{Acknowledgement}
This work is supported by MEXT Quantum Leap Flagship Program (MEXT Q-LEAP) Grant No. JPMXS0118067394.
A part of this work was performed for Council for Science, Technology and Innovation (CSTI), Cross-ministerial Strategic Innovation Promotion Program (SIP), "Photonics and Quantum Technology for Society 5.0" (Funding agency: QST).
ST is supported by CREST (Japan Science and Technology Agency) JPMJCR1671 and QunaSys Inc. YON acknowledges Takao Kobayashi for inspiring discussion to bring this project out.
YON and ST acknowledge valuable discussions with Kosuke Mitarai and Wataru Mizukami.
ST acknowledges Masato Koashi for valuable discussions.
We acknowledge IBM Q Startup program for providing access to IBM Q cloud computers which are used in our experiment.
A part of the numerical simulations in this work were done on Microsoft Azure Virtual Machines provided through the program Microsoft for Startups.

\appendix
\section{Hamiltonian for LiF \label{app: Ham LiF}}
The model Hamiltonian of a LiF molecule at a bond length $R$ under the Born-Oppenheimer approximation, $H_{\mr{LiF}}(R)$, is constructed by the following steps. (1) The state-average CASSCF method with the active space of (6 orbital, 6 electrons) is carried out by adopting the aug-ccppvdz basis set. The state-average is taken for two lowest $^1\Sigma^+$ states. (2) We pick two lowest $\sigma, \sigma^*$ molecular orbitals from the six optimized orbitals of CASSCF and construct a fermionic Hamiltonian by using them. (3) The parity mapping method~\cite{Bravyi2002, Seeley2012} is employed to map the fermionic Hamiltonian to the qubit Hamiltonian; the number of qubits required is four at this point.
Two qubits among the four are frozen from the symmetry constraints for the number of electrons and the number of total z-component of the spin~\cite{bravyi2017tapering}, which finally results in the two-qubit Hamiltonian.
The construction of the Hamiltonian is processed by PySCF~\cite{https://doi.org/10.1002/wcms.1340} and OpenFermion~\cite{McClean_2020}.

\section{Interpolation of the results to perform TSH \label{app: interpolation}}
As described in Sec.~\ref{sec: experiment}, we interpolate the values of the (approximate) eigenenergies $\tilde{E}_0, \tilde{E}_1$ and the 1-NAC $\tilde{d}_{0,1}$ evaluated at the finite number of points and supply them to the programming code for the TSH molecular dynamics simulation.
Sixty-six points in the range of $2$ to $15\AA$ are used for evaluation, and we perform the cubic spline interpolation for them implemented in Scipy, a numerical library in Python~\cite{2020NatMe..17..261V}.

\section{Cost analysis of our algorithms \label{app: cost ours}}
\subsection{Cost of our algorithm for 1-NAC}
To estimate the number of measurements to evaluate the 1-NAC with our method, let us consider the error in estimating expectation values of the Hamiltonian $H(\vec{R})$ and $\frac{\partial H(\vec{R})}{\partial R}$. 
We write $H(\vec{R}) = \sum_{i=1}^{N_H} h_i(\vec{R}) P_i$ and $\partial H(\vec{R})/\partial R_I = \sum_{i=1}^{N_H} \partial h_i(\vec{R})/\partial R_I P_i$ as the same in the main text.
The Hoeffding's inequality~\cite{Hoeffding} implies that an expectation value $\braket{P_i}$ can be estimated within the precision $\epsilon_{P}$ with high probability $1-\delta$ by measuring $P_i$ for $O(\ln(1/\delta)/\epsilon^2_{P})$ times.
When we perform $O(\ln(1/\delta)/\epsilon^2_{P})$ measurements for estimating each $\braket{P_i}$, the total error in estimating the expectation value of $H(\vec{R})$ is
\begin{equation}
    \left| \widetilde{\braket{H(\vec{R})}} - \braket{H(\vec{R})} \right| \leq \epsilon_{P} \sum_{i=1}^{N_H} |h_i| = \epsilon_{P} \|H\|,
    \label{eq: error hamiltonian}
\end{equation}
where $\widetilde{\cdots}$ is an estimated value of $\cdots$ and $\|H\| = \sum_{i=1}^{N_H} |h_i|$.
In the same way, the error of $\langle \frac{\partial H(\vec{R})}{\partial R_I} \rangle$ is given as
\begin{equation}
    \left| \widetilde{\Braket{\frac{\partial H(\vec{R})}{\partial R_I}}} - \Braket{\frac{\partial H(\vec{R})}{\partial R_I}} \right| \leq \epsilon_{P} \left\|\frac{\partial H}{\partial R_I}\right\|,
    \label{eq: error deriv. hamiltonian}
\end{equation}
where $\left\| \partial H / \partial R_I \right\| = \sum_{i=1}^{N_H} \left| \partial h_i / \partial R_I \right|$.

By using Eqs.~\eqref{eq: error hamiltonian}\eqref{eq: error deriv. hamiltonian}, we can derive the total number of measurements needed to estimate the 1-NAC within error $\epsilon$.
Equation~$\eqref{eq: 1-NAC formula}$ is evaluated in our method as
\begin{equation}
 d_{k,l}^{I}  = -\frac{\bra{\varphi_{k,l}^{+}}\frac{\partial H(\vec{R})}{\partial R_I}\ket{\varphi_{k,l}^+}-\bra{\varphi_{k,l}^{-}}\frac{\partial H(\vec{R})}{\partial R_I}\ket{\varphi_{k,l}^-}}{2(\bra{\varphi_k}H(\vec{R})\ket{\varphi_k}-\bra{\varphi_l}H(\vec{R})\ket{\varphi_l})}.
\label{eq: decompose 1-NAC}
 \end{equation} 
 
When we perform $O(\ln(1/\delta)/\epsilon^{2}_P)$ measurements to estimate the expectation value of each $P_i$ appearing in $\braket{\varphi_{k,l}^\pm|\frac{\partial H(\vec{R})}{\partial R_I}|\varphi_{k,l}^\pm}, \braket{\varphi_{k(l)}|H(\vec{R})|\varphi_{k(l)}}$,
the error propagation follows that the total error is given by
 \begin{equation}
    \left| \widetilde{d_{k,l}^{I}} - d_{k,l}^{I} \right| \lesssim \frac{2\epsilon_P}{|\Delta E_{k,l}|^2}
    \left( |\Delta E_{k,l}| \left\| \frac{\partial H}{\partial R_I}\right\| + \|H\| |A_I| \right),
    \label{eq: the error of the 1-NAC}
 \end{equation}
 where $A_I, \Delta E_{k,l}$ are the numerator and denominator of Eq.~\eqref{eq: decompose 1-NAC}, respectively.
To upper bound the error of $d_{k,l}^{I}$ by $\epsilon$, it is enough to set
 \begin{equation}
    \epsilon_{P,I} = \frac{\epsilon |\Delta E_{k,l}|^2}{2 \left( |\Delta E_{k,l}| \left\| \frac{\partial H}{\partial R_I}\right\| + \|H\| |A_I| \right)}.
 \end{equation}
Recalling that the 1-NACs for all $I=1,\ldots,N_x$ are obtained by the same results of the measurements $\braket{P_i}$, we conclude that the total number of measurements to estimate all the 1-NACs within the error $\epsilon$ with probability $1-\delta$ is given by
 \begin{equation}
 \begin{split}
     &N_{\mr{total}}^\mr{1\mathchar`-NAC} \\
     &= O\left[\frac{N_H\ln(1/\delta)}{\epsilon^2 |\Delta E_{k,l}|^4 }\left( |\Delta E_{k,l}| \left\| \frac{\partial H}{\partial R_{I^*}}\right\| + \|H\| |A_{I^*}| \right)^2 \right],
 \end{split}
 \end{equation}
 where $I^* = \mr{argmax}_I \left( |\Delta E_{k,l}| \left\| \frac{\partial H}{\partial R_{I}}\right\| + \|H\| |A_{I}| \right)$.

\subsection{Cost of the finite difference method for 1-NAC}
Let us discuss the number of measurements to calculate the 1-NAC with the finite difference method based on Eq.~\eqref{eq: numerical differentiation of 1-NAC}. 
Let $s_{k,l}^{\pm, I}$ denote the overlap $\left|\braket{\psi_k|U^{\dag}(\vec{\theta}^{*}(\vec{R}))U(\vec{\theta}^{*}(\vec{R}\pm h\vec{e}_I))|\psi_l}\right|^2$ and we compute $\tau_{k,l}^{\pm, I}$ in Eq.~\eqref{eq: numerical differentiation of 1-NAC} as $\tau_{k,l}^{\pm,I} = \sqrt{s_{k,l}^{\pm,I}}$.
By using Hoeffding's inequality, we know that $O(\ln(1/\delta)/\epsilon_{s}^2)$ projective measurements for the state $U^{\dag}(\vec{\theta}^{*}(\vec{R}))U(\vec{\theta}^{*}(\vec{R}\pm h\vec{e}_I)\ket{\psi_l}$ onto the computational basis state $\ket{\psi_k}$ is required to estimate $s_{k,l}^{\pm, I}$ within error $\epsilon_s$ with probability $1-\delta$.

When the error of the overlap $s_{k,l}^{\pm, I}$ is bounded by $\epsilon_s$, or
$ \left| \widetilde{s_{k,l}^{\pm,I}} - s_{k,l}^{\pm,I} \right| \leq \epsilon_s$,
it follows that
\begin{equation}
    \left| \widetilde{\tau_{k,l}^{\pm,I}} - \tau_{k,l}^{\pm,I} \right| \lesssim \frac{\epsilon_s}{2\tau_{k,l}^{\pm,I}}.
\end{equation}
The error of the 1-NAC with the finite difference method is then be expressed as
\begin{equation}
    \left| d_{k,l}^I - \frac{\widetilde{\tau_{k,l}^{+,I}}- \widetilde{\tau_{k,l}^{-,I}}}{2h}\right|
    \leq \frac{h^2}{6}M_3^{(I)}+\frac{\epsilon_s}{2h\tau_{k,l}^{I}},
\end{equation}
where $\tau_{k,l}^{I} = \min\{ \tau_{k,l}^{+, I}, \tau_{k,l}^{-, I}\}$, $M_3^{(I)} = \max_{s\in[-h,h]} \left|\tau_{k,l}^{(I, 3)}(s) \right|$ and $\tau_{k,l}^{(I, 3)}(s) = \frac{d^3}{ds^3} \tau_{k,l}(\vec{R}, \vec{R}+s\vec{e}_I)$.
To upper bound the right-hand side by $\epsilon$, we have to choose $\epsilon_s$ as 
\begin{equation}
    \epsilon_s = 2\tau_{k,l}^{I}\left(\epsilon h - \frac{h^3}{6}M_3^{(I)}\right).
\end{equation}
Therefore, the total number of measurements needed to evaluate the 1-NAC for all $I=1,\ldots,N_x$ within the precision $\epsilon$ with the finite difference method is
\begin{equation}
    {N^{'}}_{\mr{total}}^\mr{1\mathchar`-NAC}
    = O\left[\frac{N_x \ln(1/\delta)}{T_{k,l}^2\left(\epsilon h - \frac{h^3}{6}M_3\right)^2}\right],
\end{equation}
where $T_{k,l}^2 = \min_I \tau_{k,l}^I$ and $M_3 = \max_I M_3^{(I)}$.
Moreover, to clarify the dependence of $\epsilon$, we take $h$ such that $\epsilon_s$ attains the maximum with respect to $h$ and that ${N^{'}}_{\mr{total}}^\mr{1\mathchar`-NAC}$ takes the minimum.
This is realized for $h=\sqrt{\frac{2\epsilon}{M_3}}$ and we obtain $\epsilon_{s,\mr{max}}=\frac{4\sqrt{2}}{3}T_{k,l}M_3^{-\frac{1}{2}}\epsilon^{\frac{3}{2}}$.
Under the assumption that $M_3\geq O(\epsilon)$, we have  $\epsilon_{s,\mr{max}}\leq \frac{4\sqrt{2}}{3}T_{k,l}\epsilon$.
In such case, it follows that
\begin{equation}
    {N^{'}}_{\mr{total}}^\mr{1\mathchar`-NAC}=O\left(\frac{N_x\ln(1/\delta)}{T_{k,l}^2\epsilon^2}\right).
\end{equation}

\subsection{Cost of the finite difference method for 2-NAC}
The same argument applies to the error of 2-NAC with the finite difference method based on Eq.~\eqref{eq: numerical differentiation of 2-NAC}.
When we have $\left| \widetilde{\tau_{k,l}^{\pm,I}} - \tau_{k,l}^{\pm,I} \right|\leq \epsilon_s/(2\tau_{k,l}^{\pm,I})$, the error of Eq.~\eqref{eq: numerical differentiation of 2-NAC} is given by
\begin{equation}
    \left| D_{k,l}^{I}-\frac{\widetilde{\tau_{k,l}^{+}}+ \widetilde{\tau_{k,l}^{-}}}{h^2}\right|
    \leq \frac{h^2}{12}M_4^{(I)} + \frac{\epsilon_s}{h^2 \tau_{k,l}^{I}},
\end{equation}
where $M_4^{(I)} = \max_{s\in[-h,h]} \left|\tau_{k,l}^{(I, 4)}(s) \right|$ and $\tau_{k,l}^{(I, 4)}(s) = \frac{d^4}{ds^4} \tau_{k,l}(\vec{R}, \vec{R}+s\vec{e}_I)$.
To suppress the error of the 2-NAC within $\epsilon$ with the finite difference method, we have to take $\epsilon_s$ as 
\begin{equation}
    \epsilon_s = \tau_{k,l}^{I}\left(\epsilon h^2-\frac{h^4}{12}M_4^{(I)}\right),
\end{equation}
and obtain
\begin{equation}
 {N^{'}}_{\mr{total}}^{\mr{2\mathchar`-NAC}}
 =
 O\left[\frac{N_x^2}{T_{k,l}^2\left(\epsilon h^2 - \frac{h^4}{12}M_4\right)^2}\right],
\end{equation}
where $M_4 = \max_I M_4^{(I)}$.
Similarly to the analysis for the 1-NAC, when we take $h=\sqrt{\frac{3\epsilon}{M_4}}$, we obtain $\epsilon_{s, \mr{max}}=\frac{9}{8}\frac{\epsilon^2}{M_4}$.
If $M_4\geq O(\epsilon)$, it follows $\epsilon_{s,\mr{max}}\leq \frac{9}{8}\epsilon$.
Finally, we reach 
\begin{equation}
    {N^{'}}_{\mr{total}}^{\mr{2\mathchar`-NAC}}=O\left(\frac{N_x^2\ln(1/\delta)}{T_{k,l}^2\epsilon^2}\right).
\end{equation}

\section{Numerical simulations for NACs and Berry's phase \label{app: numerical simulation}}
In this section, we demonstrate our methods for calculating the NACs, the DBOC, and Berry's phase by numerical simulations.
Regarding the NACs, we consider the different electronic states of the hydrogen molecules. For the DBOC, we also take the electronic state of the hydrogen molecules.
As for Berry's phase, we take a simple two-site spin model with a ``twist" parameter where Berry's phase is quantized.
In all the cases, numerical simulations of our method exhibit almost perfect agreement with the exact results.
In addition, we can reproduce the shift of the equilibrium distance of the hydrogen atom by adding the DBOC to the potential energy curve obtained by the VQE~\cite{Handy1986}. These results further validates our methods proposed in the main text.

\subsection{NACs of the hydrogen molecule}
In the numerical simulation of the NACs and the DBOC, the electronic Hamiltonians of the hydrogen molecules are prepared in bond lengths from $0.5\AA$ to $2.0\AA$ with the interval of $0.1\AA$. Furthermore, we arrange the electronic Hamiltonian around the equilibrium point from $0.7320\AA$ to $0.7350\AA$ fine enough to see the shift of the equilibrium distance with the interval of $0.0001\AA$.
We perform the standard Hartree-Fock calculation by employing STO-3G minimal basis set and compute the fermionic second-quantized Hamiltonian~\cite{RevModPhys.92.015003, Cao2019} with open-source libraries PySCF~\cite{https://doi.org/10.1002/wcms.1340} and OpenFermion~\cite{McClean_2020}.
The Hamiltonians are mapped to the sum of the Pauli operators (qubit Hamiltonians) by the Jordan-Wigner transformation~\cite{1928ZPhy...47..631J}.

\begin{figure}
\includegraphics[width=.45\textwidth, trim=0 0 0 20]{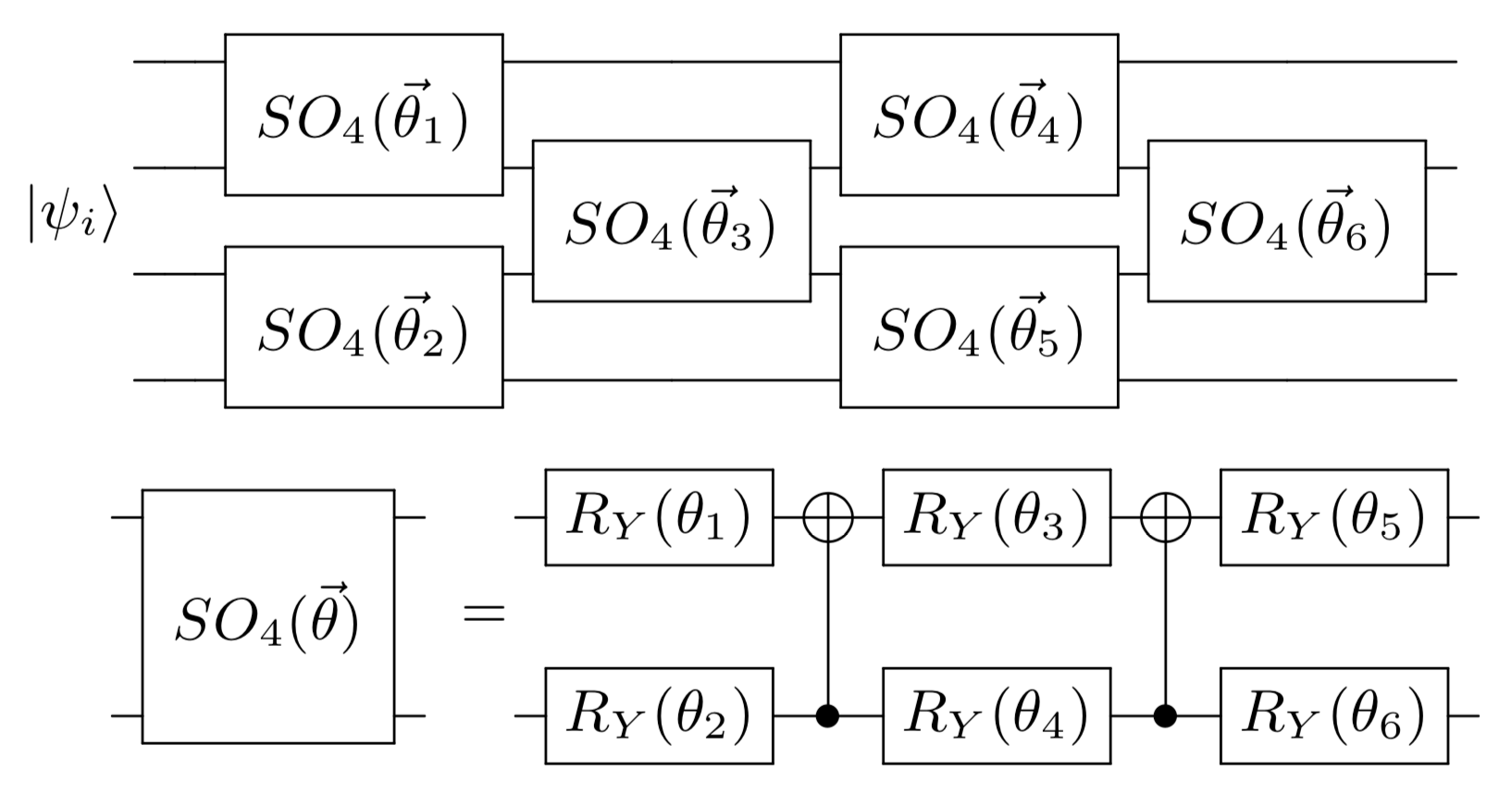}
\caption{Ansatz quantum circuit for the VQE of the hydrogen  molecule~\cite{Parrish2019}.
Each $\vec{\theta}$ has six parameters, and $R_Y(\theta)=e^{-i\frac{\theta}{2}Y}$.
The total number of parameters is 36.
\label{fig: SO4 ansatz}}
\end{figure}

The SSVQE algorithm for the qubit Hamiltonians is executed with an ansatz consisting of $SO(4)$ gates~\cite{Parrish2019} shown in Fig.~\ref{fig: SO4 ansatz}.
This ansatz gives real-valued wavefunctions for any parameters $\vec{\theta}$. 
To obtain charge-neutral and spin-singlet eigenstates, we add penalty terms containing the total particle number operator $\hat{N}$ and the total spin squared operator $\hat{S}^2$ to the Hamiltonian whose expectation value is to be minimized~\cite{kuroiwa2021}.
The cost function is
\begin{equation}
 \begin{split}
  \mathcal{L}'(\vec{\theta}) = \sum_{i=0}^{M-1}w_i\bra{\psi_i}U^{\dag}(\vec{\theta}) H' U(\vec{\theta})\ket{\psi_i}, \\
  H' = H(\vec{R}) + \beta_S \hat{S}^2 + \beta_N (\hat{N}-N_0)^2,
\end{split}
\end{equation}
where $N_0=2$ is the number of electrons and $\beta_S=\beta_N=10$ are the penalty coefficients.
We choose $M=3$ and obtain the singlet ground state for $i=0$ and another electronic state for $i=2$ which has a non-zero value of NACs between the ground state (i.e., having the same symmetry as the ground state).
The reference states and the weights are taken as $\ket{\psi_0}=\ket{0011},\ket{\psi_1}=\ket{0101}, \ket{\psi_2}=\ket{0110}$
and $w_0=3, w_1 =2, w_2=1$.
The circuit-parameters $\vec{\theta}$ are optimized by the \verb|BFGS| algorithm implemented in Scipy library~\cite{2020NatMe..17..261V}.
All simulations are run by the high-speed quantum circuit simulator Qulacs~~\cite{suzuki2020qulacs}. 

\begin{figure*}
\includegraphics[width=.52\textwidth, trim=20 0 20 20]{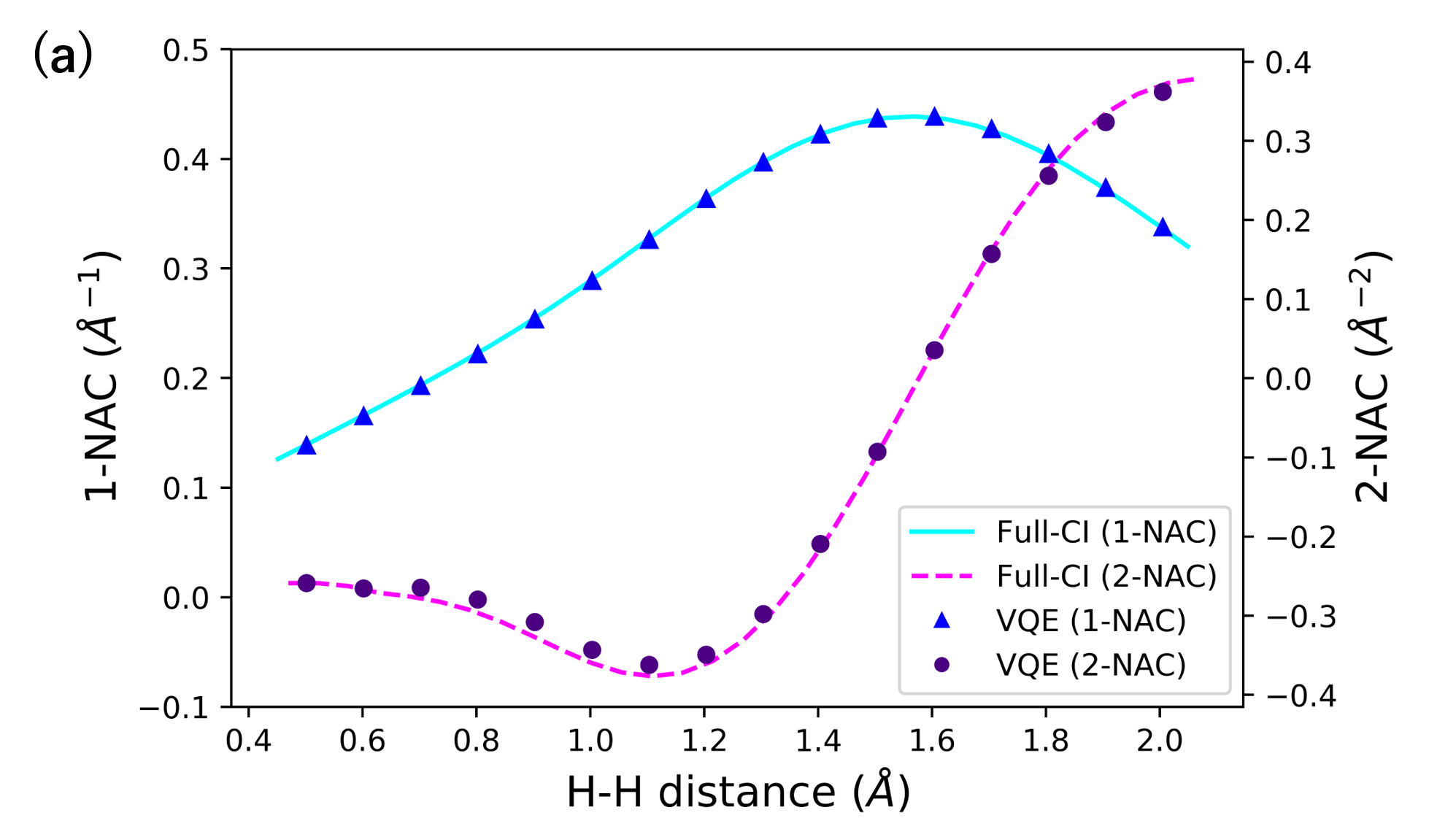}
\includegraphics[width=.47\textwidth, trim=20 0 20 20]{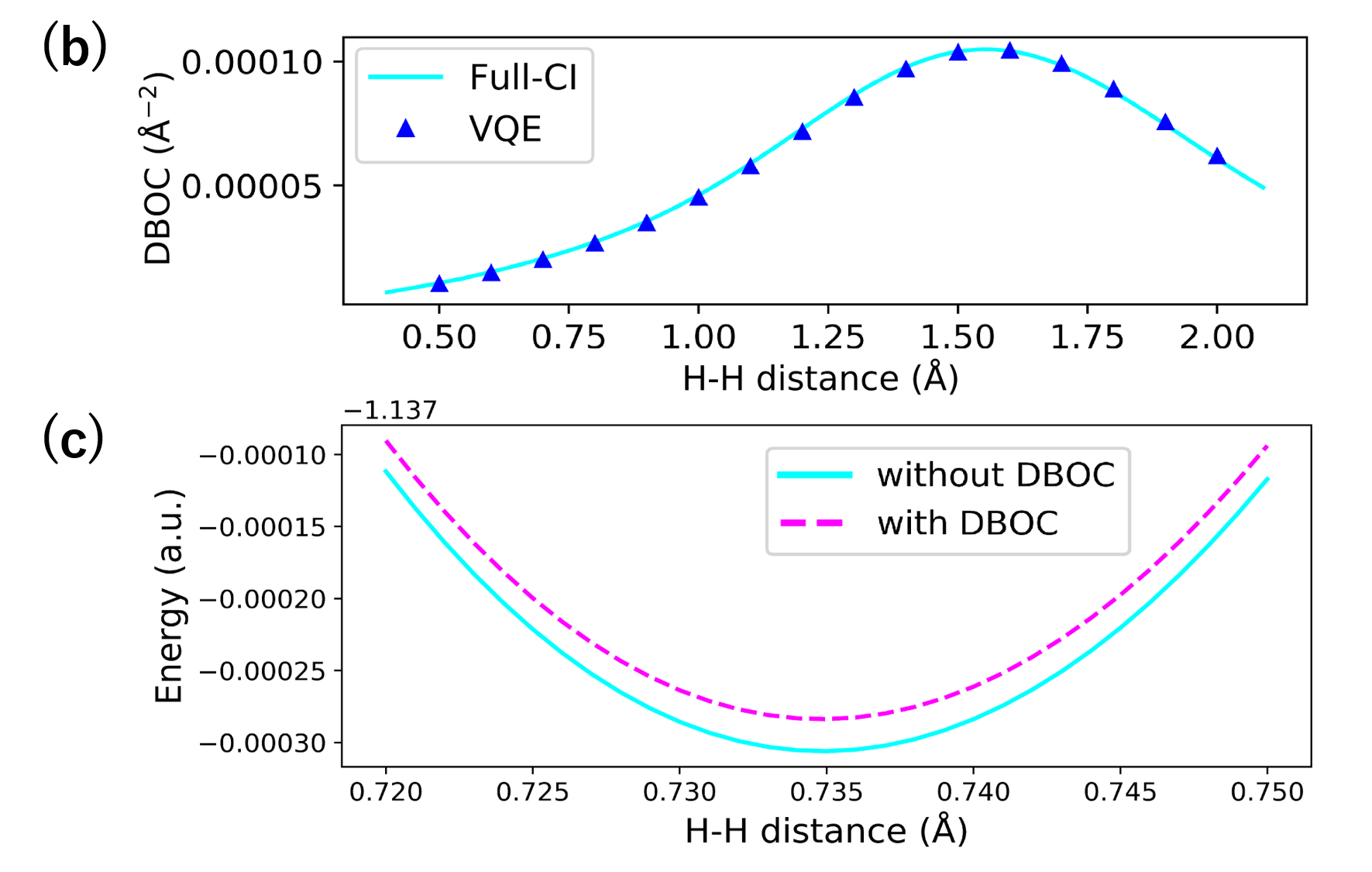}
\caption{(a) Numerical results of calculating the 1-NAC $d_{02}$ and 2-NAC $D_{02}$ between $S_0$ state and $S_2$ state of the hydrogen molecule in bond lengths from $0.5\AA$ to $2.0\AA$ with the interval of $0.1\AA$. (b) Numerical results of calculating the DBOC of $S_0$ state of the hydrogen molecule from $0.5\AA$ to $2.0\AA$ with the interval of $0.1\AA$. (c) Numerical results of calculating potential energy curves by the VQE around the equilibrium distance of the hydrogen molecule without the DBOC (solid line) and with the DBOC (dashed line) from $0.7320\AA$ to $0.7350\AA$ with the interval of $0.0001\AA$. Including the DBOC shifts the equilibrium distance from $0.7349\AA$ to $0.7348\AA$.
The NACs of ``Full-CI" are obtained by numerical differentiation of the Full-CI results.
\label{fig: result NAC}}
\end{figure*}

The results of the numerical calculation are shown in Fig.~\ref{fig: result NAC}.
We calculate the 1-NAC $d_{02}$ and 2-NAC $D_{02}$ between the ground state $i=0$ ($S_0$ state) and the excited state $i=2$ ($S_2$ state) as well as the DBOC of the ground state ($S_0$ state).
The results are in agreement with the values computed by numerical differentiation of the full configuration interaction (Full-CI) results based on the definition of the NACs (Eq.~\eqref{eq: 1-NAC def} and Eq.~\eqref{eq: 2-NAC def} in the main text). In addition, the result shown in Fig.~\ref{fig: result NAC}(c) exhibits the shift of the equilibrium distance from $0.7349\AA$ to $0.7348\AA$ by considering the DBOC based on Eq.~\eqref{eq: 2-NAC VQE} when $k=l=0$. As mentioned in the main text, the 2-NAC also has indefinite sign, thus here we determine sign by considering the continuity of the 2-NAC with respect to the nuclear coordinate.

\subsection{Berry's phase of twisted 2-spin model}
\begin{figure}
\includegraphics[width=0.45\textwidth]{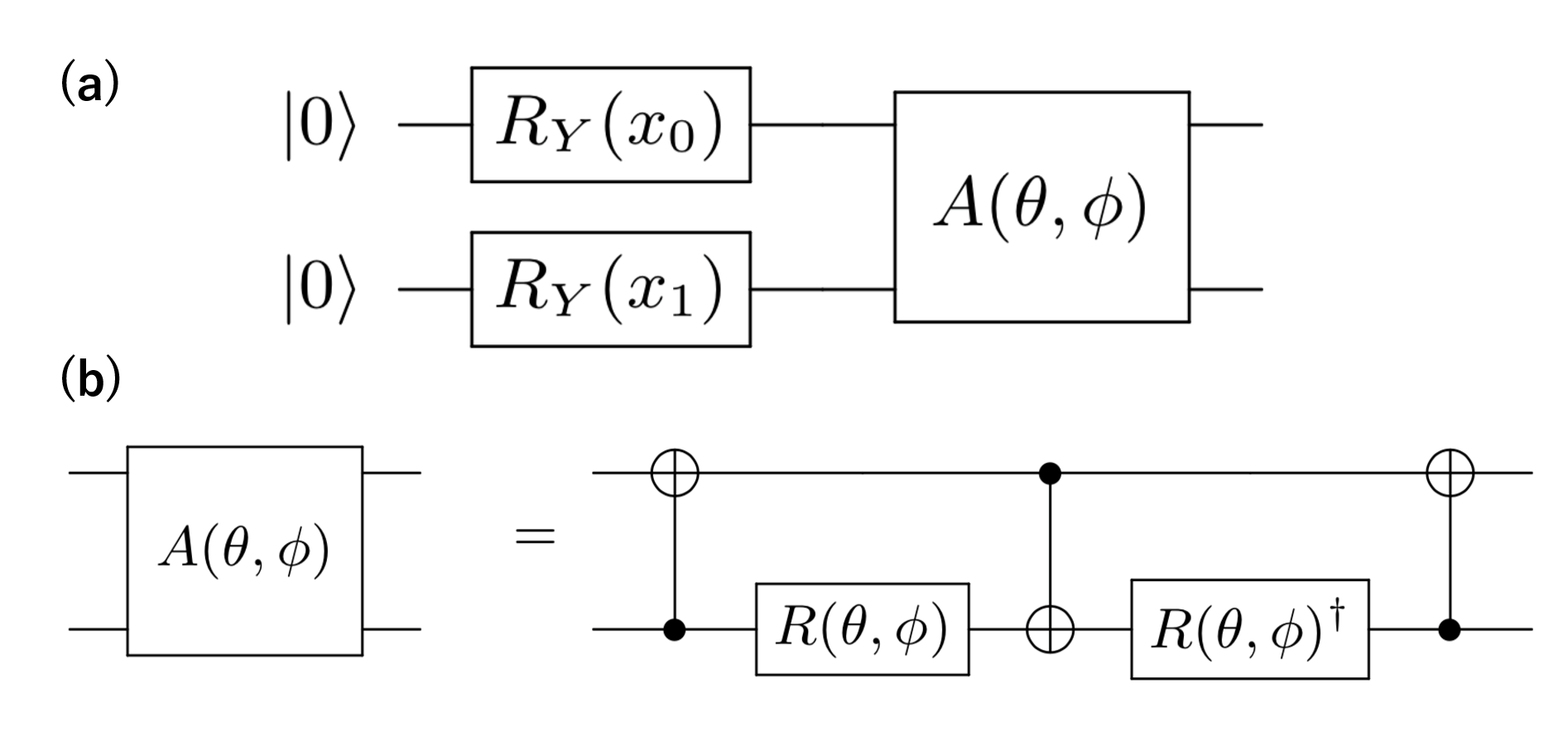}
\caption{
(a) Ansatz quantum circuit to find the ground state of Eq.~\eqref{eq: spin model}. It contains four parameters $(x_0, x_1, \theta, \phi)$ to be optimized.
(b) Definition of the particle-number-preserving gate $A(\theta, \phi)$~\cite{Gard2020}.
We define $R(\theta,\phi)$ as $R(\theta,\phi)=R_Y(\theta+\pi/2)R_Z(\phi+\pi)$, where $R_Y(\theta)=e^{i\theta Y/2}$ and $R_Z(\phi)=e^{i\phi Z/2}$.
\label{fig: ansatz for spin model} }
\end{figure}

\begin{figure}
    \includegraphics[width=8.5cm]{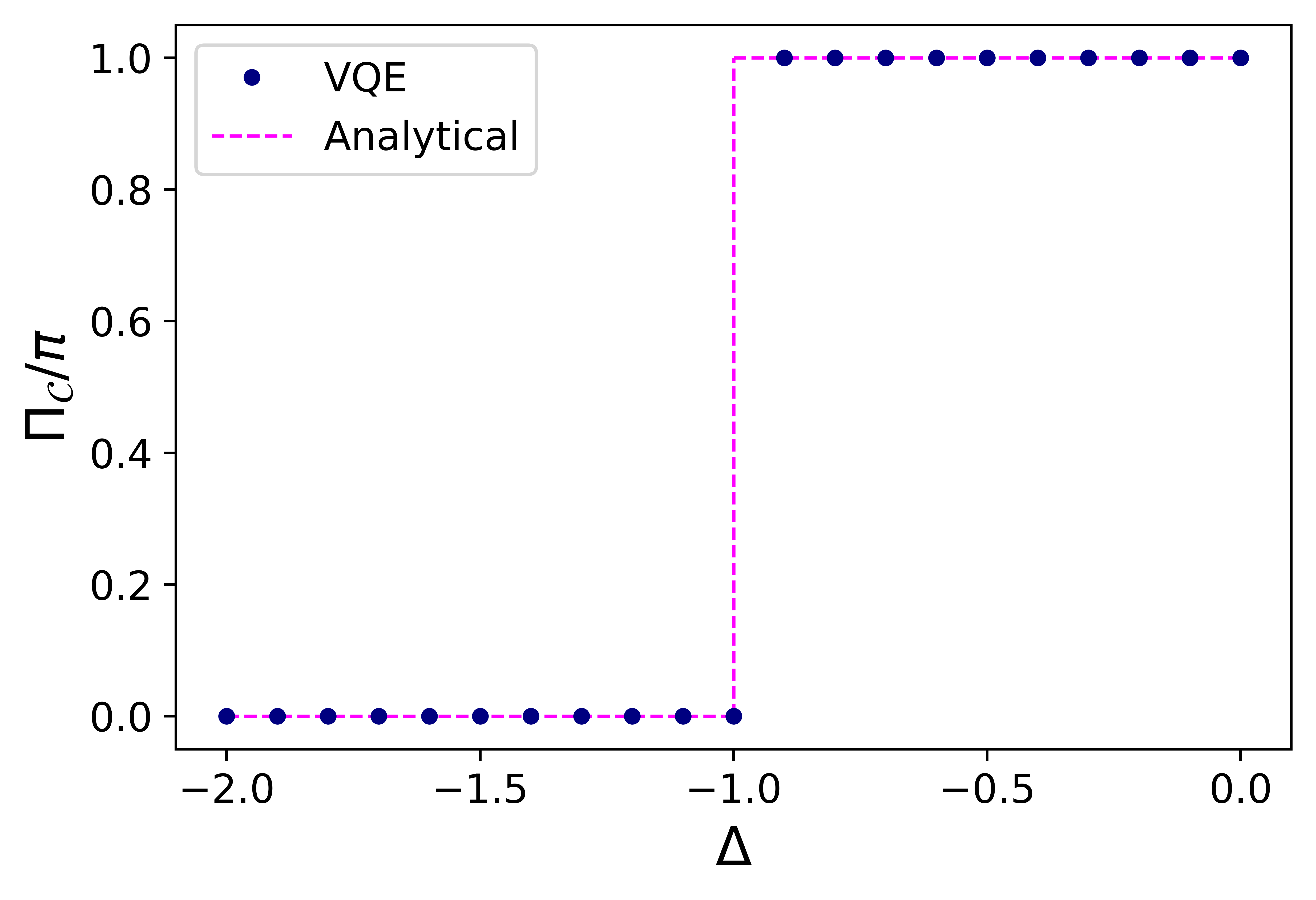}
\caption{Numerical results of Berry's phase $\Pi_\mathcal{C}$ of the model~\eqref{eq: spin model} based on Eq.~\eqref{eq: Berry's phase based on the lift} for the path $\rho=0\to2\pi$ (dots). The analytical values of the model~\eqref{eq: spin model} (dashed line). We note that the numerical result when $\Delta=-1.0$ is unstable because of the degeneracy of the ground state.
\label{fig: result Berry}}
\end{figure}

To demonstrate our method for Berry's phase, we use a two-site spin-1/2 model with a twist.
The Hamiltonian is defined as
\begin{equation}
  H_{\Delta}(\rho) = - \frac{1}{2} \left( e^{-i\rho} S_0^+ S_1^- + e^{i\rho} S_0^- S_1^+ \right) + \Delta S_0^z S_1^z,
  \label{eq: spin model}
\end{equation}
where $S_i^\pm = \frac{1}{2}(X_i \pm Y_i)$, $S_i^z = \frac{1}{2}Z_i$, and $\rho$ is a twist angle.
$\Delta$ is the parameter determines type and strength of the interaction between spins.
The ground state for $-1 < \Delta$ of this model is
\begin{equation}
 \ket{\chi_0(\rho)} = \frac{1}{\sqrt{2}} \left( \ket{01} + e^{i\rho}\ket{10} \right),
\end{equation}
while for $\Delta < -1$ it is degenerate as
\begin{equation}
 \ket{\chi_0(\rho)} = \ket{00}, \ket{11}.
\end{equation}
Since $H(\rho=0)=H(\rho=2\pi)$, we can consider Berry's phase $\Pi_\mathcal{C}$ associated to the closed path $\mathcal{C}$ from $\rho=0$ to $\rho=2\pi$.
From the exact expression of the ground state above, the analytical values of $\Pi_\mathcal{C}$ can be calculate as $\Pi_\mathcal{C}=\pi$ for $-1<\Delta$ and  $\Pi_\mathcal{C}=0$ for $\Delta < -1$.
Berry's phase $\Pi_\mathcal{C}$, in this case, is called the local $Z_2$ Berry's phase and known to detect the topological nature of the ground state of quantum many-body systems~\cite{Hatsugai2006, Kariyado2018}.

We perform the VQE for the model~\eqref{eq: spin model} with the ansatz depicted in Fig.~\ref{fig: ansatz for spin model}.
Again, the \verb|BFGS| algorithm implemented in Scipy library~\cite{2020NatMe..17..261V} is used and 
all quantum circuit simulations are run by Qulacs in the noiseless case.
We discretize the path from $\rho=0$ as $\rho=2\pi$ into $100$ points uniformly and run the VQE at each point.
The first term of Eq.~\eqref{eq: Berry's phase based on the lift} is calculated by the summation~\eqref{eq: Berry theta sum} and the phase difference $\arg(\braket{\varphi(\vec{\theta}^*_{\mathrm{s}})|\varphi(\vec{\theta}^*_{\mathrm{t}})})$ in Eq.~\eqref{eq: Berry's phase based on the lift} is evaluated by the Hadamard test in Fig.~\ref{fig: Hadamard test for berry}.
The result is shown in Fig.~\ref{fig: result Berry}.
The value of Berry's phase $\Pi_\mathcal{C}$ exhibits the sharp transition reflecting the change of the ground state.
These results illustrate the validity of our method to calculate Berry's phase based on the VQE.

\bibliography{bibliography}

\end{document}